\documentclass[12pt,draftclsnofoot,onecolumn]{IEEEtran}

\IEEEoverridecommandlockouts
\usepackage{indentfirst, flushend}
\usepackage{amssymb, bm, mathrsfs, times, amscd, amsmath, amsthm, amsfonts, bbm}
\usepackage{latexsym}
\usepackage{dsfont, slashbox}
\usepackage{graphicx}
\usepackage{subfigure}
\usepackage{color}
\usepackage{cases}

\newtheorem{proposition}{Proposition}

\newtheorem{remark}{Remark}
\usepackage{multirow}
\usepackage[sort]{cite}
\usepackage{multicol}
\usepackage[nooneline, flushleft]{caption2}
\usepackage{clrscode}
\usepackage{algorithm}
\usepackage{psfrag, stfloats, nicefrac}
\usepackage{amsmath}
\usepackage{amssymb}

\usepackage{algorithmic}
\usepackage{supertabular}
\usepackage{threeparttable}
\usepackage{makecell}
\usepackage{epstopdf}
\allowdisplaybreaks[4]

\begin{document}

\title{An Unsupervised Deep Unrolling Framework for Constrained Optimization Problems in Wireless Networks}
\author{Shiwen~He,~\IEEEmembership{Member,~IEEE}, Shaowen Xiong, Zhenyu An,~\IEEEmembership{Student Member,~IEEE}, Wei Zhang, Yongming Huang,~\IEEEmembership{Senior Member,~IEEE}, and Yaoxue Zhang,~\IEEEmembership{Senior Member,~IEEE}
\thanks{S. He, S. Xiong, and W. Zhang are with the School of Computer Science and Engineering, Central South University, Changsha 410083, China. S. He is also with the National Mobile Communications Research Laboratory, Southeast University, and the Purple Mountain Laboratories, Nanjing 210096, China. (email: \{shiwen.he.hn, shaowen.xiong, csuzwzbn\}@csu.edu.cn).}
\thanks{Z. An is with the Purple Mountain Laboratories, Nanjing 210096, China.  (email: anzhenyu@pmlabs.com.cn).}
\thanks{Y. Huang is with the National Mobile Communications Research Laboratory, School of Information Science and Engineering, Southeast University, Nanjing 210096, China. He is also with the Purple Mountain Laboratories, Nanjing 210096, China. (email: huangym@seu.edu.cn). }
\thanks{Y. Zhang is with the Department of Computer Science and Technology, Tsinghua University, Beijing 100084, China. (email: zhangyx@tsinghua.edu.cn)}
}

\maketitle
\vspace{-.6 in}

\begin{abstract}
In wireless network, the optimization problems generally have complex constraints, and are usually solved via utilizing the traditional optimization methods that have high computational complexity and need to be executed repeatedly with the change of network environments. In this paper, to overcome these shortcomings, an unsupervised deep unrolling framework based on projection gradient descent, i.e., unrolled PGD network (UPGDNet), is designed to solve a family of constrained optimization problems. The set of constraints is divided into two categories according to the coupling relations among optimization variables and the convexity of constraints. One category of constraints includes convex constraints with decoupling among optimization variables, and the other category of constraints includes non-convex or convex constraints with coupling among optimization variables. Then, the first category of constraints is directly projected onto the feasible region, while the second category of constraints is projected onto the feasible region using neural network. Finally, an unrolled sum rate maximization network~(USRMNet) is designed based on UPGDNet to solve the weighted SR maximization problem for the multiuser ultra-reliable low latency communication system. Numerical results show that USRMNet has a comparable performance with low computational complexity and an acceptable generalization ability in terms of the user distribution.
\end{abstract}


\section*{\sc \uppercase\expandafter{\romannumeral1}. Introduction}
Driven by the extensive deployment of the fifth generation~(5G) communication systems and the researches of the 6G communication technologies, various emerging wireless applications, e.g., ultra-reliable low latency communication~(uRLLC), are becoming the most innovative technical motivations, which would be expected to support various quality-of-service~(QoS) requirements~\cite{sutton2019enabling}. To satisfy the requirement of low latency, the transmission schemes not only need to exploit the network resources efficiently, but also should be executed as fast as possible~\cite{he2021survey}. However, the corresponding optimization problems are usually very complicated, which are hard to obtain their closed form solutions. In general, for these complex optimization problems, iterative optimization schemes, e.g., interior point method, should be utilized with the cost of high computational complexity. W. R. Ghanem~$et~al.$ investigated the optimal resource allocation algorithm design based on successive convex approximation~(SCA) for broad-band multiple-input single-output~(MISO) orthogonal frequency division multiplexing~(OFDMA) uRLLC systems~\cite{ghanem2020resource}. W. R Ghanem $et~al.$ firstly studied the resource allocation prolem for intelligent reflecting surface~(IRS) aided MISO OFDM uRLLC systems and proposed a suboptimal iterative optimization algorithm~\cite{ghanem2021joint}. A. A. Nasir $et~al.$ considered a downlink uRLLC system in the finite blocklength regime, and solved three different optimization problems with the objective of maximizing the users’ minimum rate using several appropriate methodologies and various convex/concave bounds~\cite{nasir2020resource}. They also proposed a particular class of conjugate beamforming for a cell free massive multiple-input multiple-output~(MIMO) downlink uRLLC system to maintain the low computational complexity~\cite{nasir2021cell}. S. He $et~al.$ focused on the beamforming design for the downlink multiuser uRLLC system~\cite{he2021beamforming}. They proposed three algorithms based on SCA to solve three different optimization problems subjected to some complicated constraints. Due to the strict requirements of latency and QoS in uRLLC scenarios, the optimization problems studied in the aforementioned works inevitably are considerable complex. Although these algorithms achieved a better performance, they face the problem of high computational complexity.

In addition to the high computational complexity of traditional optimization schemes, another issue is that the optimization scheme should be executed repeatedly for each wireless network realization, which will further decrease the efficacy of transmission schemes. To overcome these challenges, a promising way is to learn a mapping from the wireless network realization to the (sub)-optimal transmission scheme using deep neural networks~(DNNs), which benefits from the properties of universal approximation and faster inference speed~\cite{sun2018learning}. In recent years, many researchers began to use DNNs to solve the problems in wireless networks. M. Kulin~$et~al.$ studied the works on the application of DNNs in physical layer, media access control layer and network layer of wireless network~\cite{HeGNNOverview2021}. However, these works mainly use the traditional DNNs that operated in Euclidean domain, which are not suitable for wireless network with disordered communication devices and hard to exploit the non-Euclidean information in wireless network. In recent years, the rising graph neural networks~(GNNs) make up for these shortcomings faced by the traditional DNNs, especially benefit by its permutation equivariance~(PE). S. He $et~al.$ studied the application of GNNs comprehensively in wireless networks~\cite{MLwirelessLayerSurvey2021}. J. Guo considered the power control problem in multi-cell cellular networks~\cite{guo2021learning}. Specifically, this work regarded the cellular networks as a heterogeneous graph, and then proposed a heterogeneous GNN to learn the power control policy. Z. Wang $et~al.$ addressed the asynchronous decentralized wireless resource allocation problem with a novel unsupervised learning approach based on GNN~\cite{Wangemphetal2021}. M. Lee $et~al.$ analyzed and enhanced the robustness of the decentralized GNN in different wireless communication systems, making the prediction results not only accurate but also robust to transmission errors~\cite{Leeemphetal2021}. M. Eisen $et~al.$ introduced random edge GNNs~(REGNNs), which performs convolutions over random graphs in the wireless network and the REGNN-based resource allocation policies retain an important PE property that makes them amenable to transference to different networks~\cite{REGNN2020}. To train the REGNN with complex constraints, the authors proposed a learning scheme based on Lagrange primal dual thoery. Y. Shen $et~al.$ utilized GNNs to design a message passing GNN~(MPGNN) to solve the challenging radio resource management problems in wireless networks~\cite{shen2020graph}. However, the proposed MPGNN can only be used to deal with optimization problem with simple constraints.

In fact, the optimization problem in uRLLC network usually subjects to many complex constraints. That is, how to learn the mapping with complex constraints is also a challenging task for the design of transmission schemes in wireless communication systems. Recently, for given wireless network context, many researchers are focusing on solving the constrained optimization problems using DNNs. Y. Shen~\emph{et al.} proposed a learning framework for resource management to learn the optimal pruning policy in the branch-and-bound algorithm for mixed-integer nonlinear programming via imitation learning to reduce the computational complexity~\cite{LORM2020}. C. Sun~\emph{et al.} proposed a universal unsupervised deep learning framework based on Lagrange primal dual theory to solve the optimization problems with instantaneous statistic constraints in wireless communication systems~\cite{sun2020unsupervised}. J. Li $et~al.$ proposed a joint scheduling method to achieve long-term QoS tradeoff between enhanced mobile broadband~(eMBB) service and uRLLC service~\cite{Li2020}. Specifically, they jointly optimized the bandwidth allocation and overlapping positions of uRLLC users’ traffic with deep deterministic policy gradient algorithm observing channel variations and uRLLC traffic arrivals. More recently, W. Lee $et~al.$ utilized DNNs to learn the resource allocation scheme to assure the QoS in device-to-devie~(D2D) communication systems~\cite{LeeRADNN2021}. M. Alsenwi $et~al.$ studied the resource slicing problem in uRLLC and eMBB system based on reinforcement learning aiming at maximizing the eMBB data rate with complex constraints~\cite{Alsenwiemphetal2021}.

The aforementioned works solved the constrained optimization problems mainly using the pure data-driven DNNs with a poor interpretability. To overcome this defect, deep unrolling~(or unfolding) technique becomes a promising tool, which combines the advantages of model-driven algorithms and data-driven DNNs, and has been applied in various application scenarios~\cite{AlgUnroll2021}. A. Jagannath $et~al.$ reviewed the deep unfolded approaches and positioned these approaches explicitly in the context of the requirements imposed by the next generation of cellular networks~\cite{DLUnroll2021}. H. He $et~al.$ developed a model-driven DL network for MIMO detection without any constraints~\cite{he2018model}. W. Xia $et~al.$ introduced general data- and model-driven beamforming NNs for mobile communication networks subjecting to a simple power constraint~\cite{xia2020model}. Q. Hu~$et~al.$ unrolled the weighted minimum mean square error~(WMMSE) algorithm into a layer-wise structure to solve the sum rate maximization problem with simple power constraints in multiuser MIMO systems~\cite{hu2020iterative}. Similarly, WMMSE is unfolded combined with GNNs to solve the power allocation problem with simple power constraints in a single-hop Ad hoc wireless network~\cite{UWMMSE2021}. Y. Shi $et~al.$ developed an unrolled DNN framework to support grant-free massive access in IoT networks, which keeps the low computational complexity by inheriting the structure of iterative shrinkage thresholding algorithm~\cite{shi2021algorithm}. Q. Wan $et~al.$ proposed an unrolled deep learning architecture based on inverse-free variational Bayesian learning framework for MIMO detection~\cite{wan2021variational}. X. Ma $et~al.$ proposed a model-driven channel estimation and feedback learning scheme for wideband millimeter-wave massive hybrid MIMO systems~\cite{CEmmWaveUnroll2021}. Although these works achieve a better performance with lower computational complexity than that of the traditional model-driven methods, they are still lack of the ability of solving the optimization problems with complex constraints.

In this paper, we propose a universal deep unrolling framework based on projection gradient descent~(PGD), i.e., unrolled PGD network~(UPGDNet), to solve a family of constrained optimization problems in wireless communication networks. The main contributions are listed as follows
\begin{itemize}
  \item Firstly, we divide separate the constraints into two categories according to the coupling relations among optimization variables and the convexity of constraints. One category of constraints includes convex constraints with decoupling among optimization variables, and the other category of constraints includes non-convex or convex constraints with coupling among optimization variables. Then, for one category of constraints, we directly project them onto the feasible region, while the other category of constraints are projected onto the feasible region using a NN.
  \item Secondly, we propose the UPGDNet to address the problem of interest. Further, we propose a Lagrange primal dual learning framework with a multi-task loss function to train the UPGDNet stably in an unsupervised manner.
  \item Thirdly, To verify the effectiveness of the UPGDNet, we utilize it to solve the weighted sum rate maximization~(WSRMax) problem in the scenario of multiuser uRLLC with finite blocklength transmission.
  \item Finally, numerical results show that the UPGDNet can be trained efficiently using our proposed training scheme. In addition, the unrolled sum rate maximization network~(USRMNet) model consisting of UPGDNet has a comparable performance with the baseline algorithm on the basis of ensuring low computational complexity. In addition, the USRMNet model also has a acceptable generalization ability in terms of the UE distribution~\footnote{The codes to reproduce the simulation results are available on https://github.com/SoulVen/USRMNet-HWGCN.}.
\end{itemize}

The rest of this paper is organised as follows. Section II describes the family of constrained optimization problem and proposed deep unrolling framework. In Section III, we utilize the deep unrolling framework to solve the WSRMax problem in uRLLC systems. In Section IV, we present several numerical simulation results to verify the effectiveness of the UPGDNet. Finally, we conclude this paper in Section V.

\textit{Notations:}~We use lower case letters and boldface capital to denote vectors and matrices, respectively. $\mathbf{a}^{H}$ denotes the Hermitian transpose of vector $\mathbf{a}$. $|\cdot|$ and $\|\cdot\|$ denote the absolute value of a complex scalar and the Euclidean vector norm. $\mathbb{C}^T$ denotes the set of complex numbers. $\mathbb{R}_+$ denotes the set of positive numbers.


\section*{\sc \uppercase\expandafter{\romannumeral2}. Description of Problem and Unrolling Method}
In this section, we firstly illustrate a family of constrained optimization problem, which is hard to be solved using traditional optimization methods. Then, to efficiently solve the family of problems, we propose a universal framework based on PGD, i.e., UPGDNet. Finally, to train the UPGDNet efficiently and stably, we further design a learning framework based on Lagrange primal dual theory and multi-task learning.
\subsection*{A. Problem Description}
In this subsection, we consider a family of constrained optimization problem, which is formulated as follows
\begin{subequations}\label{Cachenable01}
\begin{align}
&\min\limits_{\mathbf{x}}~f\left(\mathbf{x};\bm{\theta}\right), \\
\mathrm{s.t.}~&h_i\left(\mathbf{x};\bm{\theta}\right)\leq 0,i=1,\cdots,N_h,~\label{Cachenable01b}\\
&g_j\left(\mathbf{x};\bm{\theta}\right)\leq 0,j=1,\cdots,N_g,~\label{Cachenable01c}
\end{align}
\end{subequations}
where $\mathbf{x}\in\mathbb{R}^{d}$ is the variable vector that should be optimized, $\bm{\theta}\in\mathcal{D}_{\theta}\subseteq\mathbb{R}^{N_{\theta}}$ is a vector consisting of $N_{\theta}$ environmental parameters of a realization, $\mathcal{D}_{\theta}$ is a compact set of realizations, $f(\mathbf{x}; \bm{\theta})$ is a convex or non-convex objective function. $h_i(\mathbf{x})\leq0,i=1,\cdots,N_h$, and $g_j(\mathbf{x})\leq0, j=1,\cdots,N_g$, belong to constraint sets $\mathcal{C}_1$ and $\mathcal{C}_2$, respectively, where $\mathcal{C}_1$ is a set of convex constraints with decoupling among optimization variables, and $\mathcal{C}_2$ is a set of non-convex or convex constraints with coupling among optimization variables. We further assume that $f\left(\mathbf{x};\bm{\theta}\right)$, $h_i\left(\mathbf{x};\bm{\theta}\right)$, and $g_j\left(\mathbf{x};\bm{\theta}\right)$ are differentiable with respect to $\mathbf{x}$.

In some application scenarios, the constraints in constraint sets $\mathcal{C}_1$ and $\mathcal{C}_2$ may be unwieldy, which makes the optimization problem difficult to be solved directly. In the existing literature, problem~\eqref{Cachenable01} is generally solved iteratively via traditional optimization methods with high computational overhead. To deal with these challenges, an effective way is to find a mapping between $\mathbf{x}^{*}$ and $\bm{\theta}$, where $\mathbf{x}^*$ is the (sub)-optimal solution to problem~\eqref{Cachenable01}. A promising way generating the mapping $\phi\left(\bm{\theta}\right):\bm{\theta}\rightarrow\mathbf{x}^*$ is to utilize the NNs, which benefits the universal approximation property of NNs. Specifically, we rewrite problem~\eqref{Cachenable01} as follows
\begin{subequations}\label{Cachenable02}
\begin{align}
&\min\limits_{\phi(\bm{\theta})}~\mathbb{E}_{\bm{\theta}}\left\{f\left(\phi\left(\bm{\theta}\right);\bm{\theta}\right)\right\}=\int_{\bm{\theta}}f\left(\phi\left(\bm{\theta}\right);\bm{\theta}\right)p\left(\bm{\theta}\right)\mathrm{d}\bm{\theta}, ~\label{Cachenable02a}\\
\mathrm{s.t.}~&h_i(\phi\left(\bm{\theta}\right);\bm{\theta})\leq 0,\forall \bm{\theta}\in\mathcal{D}_{\theta}, i=1,\cdots,N_h,~\label{Cachenable02b}\\
&g_j(\phi\left(\bm{\theta}\right);\bm{\theta})\leq 0, \forall \bm{\theta}\in\mathcal{D}_{\theta}, j=1,\cdots,N_g,~\label{Cachenable02c}
\end{align}
\end{subequations}
where $\phi(\bm{\theta})$ is optimized to minimize the expectation of the objective function in problem~\eqref{Cachenable01}. $p\left(\bm{\theta}\right)$ is the weight coefficient of $\bm{\theta}$. How to learn the mapping $\phi\left(\bm{\theta}\right)$ which simultaneously satisfies constraints~\eqref{Cachenable02b} and~\eqref{Cachenable02c} is the main difficulty for solving problem~\eqref{Cachenable02}.

\subsection*{B. Unrolled Projection Gradient Descent Network}
In this subsection, we focus on designing a mappping $\phi\left(\bm{\theta}\right)$ and a learning framework to obtain the mappping $\phi\left(\bm{\theta}\right)$ such that problem~\eqref{Cachenable02} is solved. Specifically, due to the constraint set $\mathcal{C}_2$ is hard to be handled directly, we move constraint~\eqref{Cachenable02c} into the objective function by introducing Lagrangian multiplier vector $\bm{\lambda}=[\lambda_1,\dots,\lambda_{N_g}]^T\in\mathbb{R}^{N_g}$ to satisfy constraint set $\mathcal{C}_2$. Accordingly, the partial primal-dual problem is formulated as
\begin{subequations}\label{Cachenable05}
\begin{align}
&\max\limits_{\bm{\lambda}}\min\limits_{\phi\left(\bm{\theta}\right)}~\mathbb{E}_{\bm{\theta}}\left\{f\left(\phi\left(\bm{\theta}\right);\bm{\theta}\right)+\sum\limits_{j=1}^{N_g}\lambda_jg_j\left(\phi\left(\bm{\theta}\right);\bm{\theta}\right)\right\}, ~\label{Cachenable05a}\\
\mathrm{s.t.}~&h_i(\phi\left(\bm{\theta}\right);\bm{\theta})\leq 0,\forall \bm{\theta}\in\mathcal{D}_{\theta}, i=1,\cdots,N_h,~\label{Cachenable05b}\\
&\lambda_i \geq 0, j=1,\cdots,N_g.~\label{Cachenable05c}
\end{align}
\end{subequations}
To solve problem efficiently~\eqref{Cachenable05}, for a given initialization solution $\mathbf{x}_0\in\mathbb{R}^{d}$ for each realization $\bm{\theta}$, a preliminary solution $\hat{\mathbf{x}}$ is firstly obtained via gradient descent method, i.e., $\hat{\mathbf{x}} = \mathbf{x}_0 - \eta\nabla_{x}f\left(\mathbf{x}_0\right)$, where $\eta\geq0$ and $\nabla_{x}f\left(\mathbf{x}\right)$ denote the gradient descent step-size and the gradient of $f\left(\mathbf{x}\right)$ \text{w.r.t.} $\mathbf{x}$, respectively. Then, a perturbation vector $\bar{\mathbf{x}}\in\mathbb{R}^{d}$ is added to the preliminary solution $\hat{\mathbf{x}}$ aiming to improve the objective value while the constraint set $\mathcal{C}_2$ is satisfied. Consequently, we obtain an intermediate solution $\tilde{\mathbf{x}}=\hat{\mathbf{x}}+\bar{\mathbf{x}}$. Finally, $\tilde{\mathbf{x}}$ is projected to the space generated by constraint set $\mathcal{C}_1$, i.e.,
\begin{equation}\label{Cachenable03}
\begin{aligned}
&\mathbf{x}^* = \Pi_{C_{N_h}}\cdots\Pi_{C_2}\Pi_{C_1}\tilde{\mathbf{x}},
\end{aligned}
\end{equation}
where $C_i$ is the feasible region of the $i$-th constraint in constraint set $\mathcal{C}_1$, i.e., $C_i\triangleq\{\mathbf{x}:h_i\left(\mathbf{x};\bm{\theta}\right)\leq0\}$, $i=1,\cdots,N_h$. $\Pi_{C_i}$ is the projection onto convex set~(POCS) operation, i.e., projecting the optimization variable vector $\mathbf{x}$ onto a convex set directly, defined as
\begin{equation}\label{Cachenable04}
\begin{aligned}
&\Pi_{\left\{\mathbf{x}:\hbar(\mathbf{x})\leq\xi\right\}}\mathbf{x}=\left\{
\begin{aligned}
&\mathbf{x},~\text{if}~\hbar(\mathbf{x})\leq\xi, \\
&\mathbf{x}+\frac{\xi-\hbar(\mathbf{x})}{\|\nabla \hbar(\mathbf{x})\|^2}\nabla_{x} \hbar(\mathbf{x}),~\text{if}~\hbar(\mathbf{x})>\xi,
\end{aligned}
\right.
\end{aligned}
\end{equation}
where $\hbar(\mathbf{x}):\mathbb{R}^d\rightarrow\mathbb{R}$. The POCS operation assures the feasibility of constraints in constraint set $\mathcal{C}_1$.

\begin{figure}[t]
\renewcommand{\captionfont}{\footnotesize}
\renewcommand*\captionlabeldelim{.}
	\centering
	\captionstyle{flushleft}
	\onelinecaptionstrue
	\includegraphics[width=0.8\columnwidth,keepaspectratio]{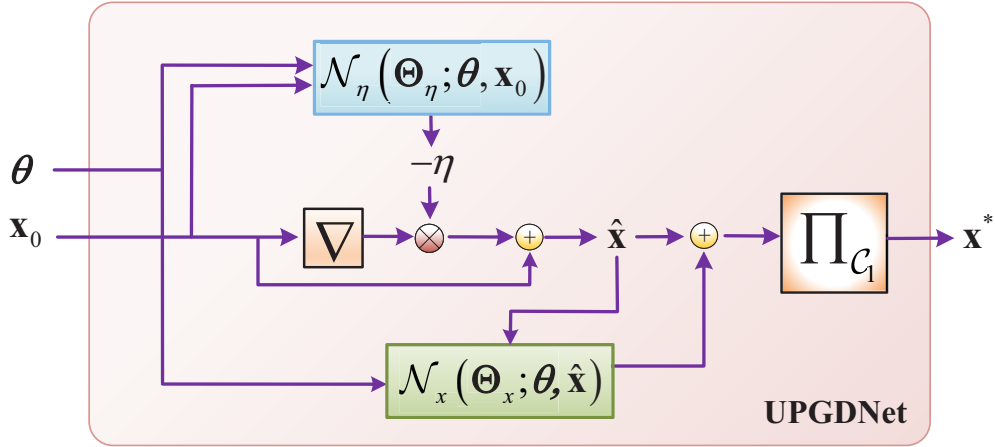}
	\caption{Illustration of the UPGDNet.}
	\label{PGDNet}
\end{figure}

In general, choosing reasonable $\eta$ and $\bar{\mathbf{x}}$ is helpful to project $\hat{\mathbf{x}}$ onto $\Omega\triangleq\{\mathbf{x}:g_i\left(\mathbf{x};\bm{\theta}\right)\leq0,i=1,\cdots,N_g\}$ and obtain the (sub)-optimal solution $\mathbf{x}^*$. However, it is not hard to find that determining the specific values of $\eta$ and $\bar{\mathbf{x}}$ is difficult, especially $\bar{\mathbf{x}}$. To overcome these difficulties, two NNs $\mathcal{N}_{\eta}\left(\mathbf{\Theta}_{\eta};\bm{\theta},\mathbf{x}_0\right)$ and $\mathcal{N}_{x}\left(\mathbf{\Theta}_{x};\bm{\theta},\hat{\mathbf{x}}\right)$ are designed to look up the appropriate $\eta$ and $\bar{\mathbf{x}}$ for each realization $\bm{\theta}$, respectively, where $\mathbf{\Theta}_{\eta}$ and $\mathbf{\Theta}_{x}$ are learnable parameter sets. The whole procedure for obtaining $\mathbf{x}^*$ is summarized in~\textbf{Algorithm}~\ref{alg:algorithm1}, namely, UPGDNet. Fig.~\ref{PGDNet} illustrates the structure of the UPGDNet, which only needs to execute PGD once to achieve the (sub)-optimal solution $\mathbf{x}^*$.

\begin{algorithm}[t]
\caption{Unrolled PGD Network~(UPGDNet)}
\label{alg:algorithm1}
\begin{algorithmic}[1]
\STATE \textbf{Input}: $\bm{\theta}$, $\mathbf{x}_0$ and other necessary parameters.
\STATE Obtain gradient descent step-size:\\~~~~$\eta\leftarrow \mathcal{N}_{\eta}\left(\bm{\Theta}_{\eta};\bm{\theta},\mathbf{x}_0\right)$.
\STATE Preliminary estimation:~$\hat{\mathbf{x}} = \mathbf{x}_0 - \eta\nabla_{x}f\left(\mathbf{x}_0\right)$.
\STATE PGD operation:\\
\STATE~~~~$\bar{\mathbf{x}}\leftarrow\mathcal{N}_{x}\left(\bm{\Theta}_x;\bm{\theta}, \hat{\mathbf{x}}\right)$, \\
\STATE~~~~$\tilde{\mathbf{x}}=\hat{\mathbf{x}}+\bar{\mathbf{x}}$, \\
\STATE~~~~$\mathbf{x}^*=\Pi_{C_{N_h}}\cdots\Pi_{C_2}\Pi_{C_1}{\tilde{\mathbf{x}}}$. 
\STATE \textbf{Output}: $\mathbf{x}^*$.
\end{algorithmic}
\end{algorithm}

In what follows, we propose a Lagrange primal dual learning framework based on problem~\eqref{Cachenable05} to train the UPGDNet, which is summarized in~\textbf{Algorithm}~\ref{alg:algorithm2}. In case of generating the ground truth and gaining performance as much as possible for each realization $\bm{\theta}$, we prefer to train the UPGDNet in an end-to-end unsupervised manner instead of supervised manner. In line 3 of \textbf{Algorithm}~\ref{alg:algorithm2}, the UPGDNet will be trained in mini-batch manner by minimizing the following loss function~\footnote{In this work, the UPGDNet is trained using Adam optimizer~\cite{kingma2014adam}. In addition, we do not consider the violation of constraint set $\mathcal{C}_1$, i.e., Eq.~\eqref{Cachenable05b}, in loss function~\eqref{joint} as it can be always satisfied by POCS operation.}
\begin{equation}\label{joint}
\begin{aligned}
&\mathcal{L}^{(\ell)}\left(\mathbf{\Theta}_{\eta},\mathbf{\Theta}_{x},\bm{\lambda}, \mathbf{s}\right) = e^{-\frac{1}{2}s_1}\mathcal{L}_1\left(\mathbf{\Theta}_{\eta},\mathbf{\Theta}_{x}\right)
+e^{-\frac{1}{2}s_2}\mathcal{L}_2\left(\mathbf{\Theta}_{\eta},\mathbf{\Theta}_{x},\bm{\lambda}\right)
+e^{\frac{1}{2}s_1}+e^{\frac{1}{2}s_2},
\end{aligned}
\end{equation}
where $\mathcal{L}_1\left(\mathbf{\Theta}_{\eta},\mathbf{\Theta}_{x}\right)=\mathbb{E}_{\bm{\theta}}\left\{f\left(\mathbf{x}_{\ell};\bm{\theta}\right)\right\}$, $\mathcal{L}_2\left(\mathbf{\Theta}_{\eta},\mathbf{\Theta}_{x},\bm{\lambda}\right)=\mathbb{E}_{\bm{\theta}}\left\{\sum_{j=1}^{N_g}\lambda_j^{\left(\ell-1\right)}g_j\left(\mathbf{x}_{\ell};\bm{\theta}\right)\right\}$, and $\mathbf{s}=[s_1, s_2]^T\in\mathbb{R}^{2\times 1}$ is learnable scale vector. $\mathbf{x}_{\ell}$ denotes the output solution in the $\ell$-th iteration. Equation~\eqref{joint} is a multi-task objective function, which is designed on the basis of the work in~\cite{kendall2018multi} to guarantee the stable training of UPGDNet. When we train the UPGDNet using loss function~\eqref{joint}, the learnable parameter set $\left\{\mathbf{\Theta}_{x}, \mathbf{\Theta}_{\eta}\right\}$, and $\mathbf{s}$ are updated based on the gradient descent method with step-size $\varepsilon_{\Theta}>0, \varepsilon_{s}>0$, respectively. The loss function~\eqref{joint} will converges towards to zero due to the items of the loss function~\eqref{joint} are all nonnegative, which guarantees the stable training of UPGDNet. After one round of the UPGDNet training, $\bm{\lambda}$ is updated based on gradient ascent with update step-size $\varepsilon_{\lambda}>0$, which is shown in line 10 of \textbf{Algorithm}~\ref{alg:algorithm2}~\footnote{The update step-sizes $\varepsilon_{\Theta}$, $\varepsilon_{s}$ and $\varepsilon_{\lambda}$ are hyper-parameters and they are manually set.}. $\left[x\right]^+$ denotes the function $\max\left(x, 0\right)$, which guarantees $\bm{\lambda}>0$ and indicates that $\bm{\lambda}$ is only updated when constraints in $\mathcal{C}_2$ are violated. Once the UPGDNet is trained properly, we can utilize the UPGDNet, i.e., \textbf{Algorithm}~\ref{alg:algorithm1}, to solve problem~\eqref{Cachenable02} directly, where the mapping $\phi\left(\bm{\theta}\right)$ in problem~\eqref{Cachenable02} is approximated by the UPGDNet, i.e., $\phi\left(\bm{\theta}\right)\triangleq\text{UPGDNet}$.

\begin{algorithm}[t]
\caption{The learning framework of the UPGDNet}
\label{alg:algorithm2}
\textbf{Input}: Step-sizes $\varepsilon_{\Theta}$, $\varepsilon_{s}$, and $\varepsilon_{\lambda}$, realization dataset $\mathcal{D}_{\theta}$, initialization solution set $\{\mathbf{x}_0^{(i)}\}_{i=1}^{|\mathcal{D}_{\theta}|}$, initialization Lagrangian multiplier vector $\bm{\lambda}^{(0)}=\bm{0}$, initialization scale vector $\mathbf{s}$, and other necessary parameters.
\begin{algorithmic}[1]
\STATE \textbf{for} $\ell = 1, 2, ...$ \textbf{do}
\STATE ~~~~\textbf{for each} $\mathcal{B}\leftarrow minibatch\left(\mathcal{D}_{\theta}\right)~of~size~b$ \textbf{do}
\STATE ~~~~~~~~Train the UPGDNet with $\varepsilon_{x}$, $\mathbf{\Theta}_{\eta}^{(\ell-1)}$, $\mathbf{\Theta}_{x}^{(\ell-1)}$, and $\mathbf{s}$ based on $\mathcal{B}$.
\STATE ~~~~~~~~~$\eta^{(\ell)}\leftarrow\mathcal{N}_{\eta}\left(\mathbf{\Theta}_{\eta}^{(\ell-1)};\bm{\theta},\mathbf{x}_0\right)$,
\STATE ~~~~~~~~~$\hat{\mathbf{x}}_{\ell}=\mathbf{x}_0-\eta_{(\ell)}\nabla_xf\left(\mathbf{x}_0\right)$,
\STATE ~~~~~~~~~$\bar{\mathbf{x}}_{\ell}\leftarrow\mathcal{N}_x\left(\mathbf{\Theta}_x^{(\ell-1)};\bm{\theta},\hat{\mathbf{x}}_{\ell}\right)$,
\STATE ~~~~~~~~~$\tilde{\mathbf{x}}_{\ell}=\hat{\mathbf{x}}_{\ell}+\bar{\mathbf{x}}_{\ell}$,
\STATE ~~~~~~~~~$\mathbf{x}_{\ell}=\Pi_{C_{N_h}}\cdots\Pi_{C_2}\Pi_{C_1}\tilde{\mathbf{x}}_{\ell}$.
\STATE \quad\quad~ Update $\mathbf{\Theta}_{\eta}, \mathbf{\Theta}_x$, and $\mathbf{s}$ according to the backward propagation of loss function~\eqref{joint}.
\STATE ~~~~~~~~Update the Lagrangian multipliers using $\mathcal{B}$:\\
~~~~~~~~$\lambda_i^{(\ell+1)}=\lambda_i^{(\ell)}+\varepsilon_{\lambda}\left[\mathbb{E}_{\bm{\theta}}\left\{g_i\left(\mathbf{x}_{\ell};\bm{\theta}\right)\right\}\right]^+,i=1,\cdots,N_g$.
\STATE ~~~~~\textbf{end for}
\STATE \textbf{end for}
\end{algorithmic}
\end{algorithm}

\section*{\sc \uppercase\expandafter{\romannumeral3}. Beamforming Design for Multiuser uRLLC system}
In this section, to validate the effectiveness of the proposed UPGDNet, we focus on solving the WSRMax problem for the downlink multiuser uRLLC system. We first introduce simply the WSRMax problem to be solved. Then, we focus on designing a model, i.e, USRMNet, based on the UPGDNet, to solve the WSRMax problem. Finally, we demonstrate the USRMNet model is permutation equivariant.
\subsection*{A. Weighted Sum Rate Maximization Problem}
In this subsection, we would like to solve the WSRMax problem, i.e., the investigated problem~(5) in~\cite{he2021beamforming} using UPGDNet. According to the \textbf{Lemma 1} in~\cite{he2021beamforming}, the WSRMax problem is solved using the uplink-downlink duality theory. For ease of notation, let $q_k$ be the transmitting power of the $k$-th user~(UE), where $k\in\mathcal{K}\triangleq\{1,\cdots,K\}$ with $K$ being the total number of UEs. Thus, the dual optimization problem is formulated as
\begin{subequations}\label{Cachenable07}
\begin{align}
&\max\limits_{\{\mathbf{w}_k, q_k\}}~\sum\limits_{k\in\mathcal{K}}\alpha_k\overleftarrow{R}\left(\overleftarrow{\gamma}_k\right), \label{Cachenable07a}\\
\mathrm{s.t.}~&\sum\limits_{k\in\mathcal{K}}q_k\leq P, \|\mathbf{w}_k\|_2^2=1, \forall k\in\mathcal{K}, \\
&\frac{D}{n}\mathrm{ln}(2)\leq \overleftarrow{R}\left(\overleftarrow{\gamma}_k\right), \forall k\in\mathcal{K},
\end{align}
\end{subequations}
where $\mathbf{w}_{k}\in\mathbb{C}^{N_{\mathrm{t}}\times 1}$ denotes the normalized beamforming vector used at base station~(BS) for  the $k$-th UE. $N_\mathrm{t}$ is the number of antennas equipped at BS. $\alpha_k$ denotes the prior of the $k$-th UE. $\overleftarrow{R}\left(\overleftarrow{\gamma}_k\right)=\ln\left(1+\overleftarrow{\gamma}_k\right)-\vartheta\sqrt{V\left(\overleftarrow{\gamma}_k\right)}$, where $\vartheta=\frac{Q^{-1}\left(\epsilon\right)}{\sqrt{n}}$ with $\epsilon$ being a desirable decoding error probability. $n$ is the finite blocklength, $D$ is the number of transmitting data bits, and $Q^{-1}\left(\cdot\right)$ being the inverse of Gaussian Q-function. $V\left(\overleftarrow{\gamma}\right)=1-\frac{1}{\left(1+\overleftarrow{\gamma}\right)^{2}}$. $P$ is the maximum allowable power constraint. $\overleftarrow{\gamma}_{k}$ is given by
\begin{equation}\label{URLLC02}
\overleftarrow{\gamma}_k=\frac{q_{k}\left|\overline{\mathbf{h}}_{k}^{H}\mathbf{w}_{k}\right|^{2}}
{\sum\limits_{l\neq k}q_{l}\left|\overline{\mathbf{h}}_{l}^{H}\mathbf{w}_{k}\right|^{2}+1},
\end{equation}
where $\overline{\mathbf{h}}_{k}=\frac{\mathbf{h}_{k}}{\sigma_{k}}$, $\mathbf{h}_{k}\in\mathbb{C}^{N_{\mathrm{t}}\times 1}$ represents the slow time-varying channel coefficient between the BS and the $k$-th UE. $\sigma_k$ is the Gaussian variance of the $k$-th UE.

Given uplink transmitting power $\{q_k\}$, the optimal solution of $\mathbf{w}_k$ for maximizing $\overleftarrow{R}\left(\overleftarrow{\gamma}_k\right)$ is the minimum mean square error receiver, i.e.,
\begin{equation}\label{Cachenable08}
\mathbf{w}_k^*=\frac{\left(\mathbf{I}_{N_{\mathrm{t}}}+\sum\limits_{k\in\mathcal{K}}q_k\overline{\mathbf{h}}_k\overline{\mathbf{h}}_k^H\right)^{-1}\overline{\mathbf{h}}_k}{\left|\left|\left(\mathbf{I}_{N_{\mathrm{t}}}+\sum\limits_{k\in\mathcal{K}}q_k\overline{\mathbf{h}}_k\overline{\mathbf{h}}_k^H\right)^{-1}\overline{\mathbf{h}}_k\right|\right|},
\end{equation}
where $\mathbf{I}_{N_{\mathrm{t}}}$ denotes $N_{\mathrm{t}}$-by-$N_{\mathrm{t}}$ identity matrix.
Given beamforming vector $\{\mathbf{w}_k\}$, the uplink power allocation problem~\eqref{Cachenable07} is reformulated as
\begin{subequations}\label{Cachenable09}
\begin{align}
&\max\sum\limits_{k\in\mathcal{K}} \alpha_k\left(\mathrm{ln}\left(1+\varphi_k\right)-\vartheta\theta_k\right), \label{Cachenable09a}\\
\mathrm{s.t.}~&\nu_3\leq\varphi_k, \forall k\in\mathcal{K}, \label{Cachenable09b}\\
&\phi_k\leq\tilde{\gamma}_k, \forall k\in\mathcal{K}, \label{Cachenable09c} \\
&V(\nu_3)-\psi_k\leq0, \forall k\in\mathcal{K}, \label{Cachenable09d}\\
&\psi_k\leq V(\tilde{\gamma}_k), \forall k\in\mathcal{K}, \label{Cachenable09e} \\
&\sum_{k\in\mathcal{K}}q_k\leq P, \label{Cachenable09f} \\
&0\leq q_k, \forall k\in\mathcal{K}. \label{Cachenable09g} \\
&\varphi_k\leq\overleftarrow{\gamma}_k, \forall k\in\mathcal{K}, \label{Cachenable09h}\\
&\overleftarrow{\gamma}_k\leq\phi_k, \forall k\in\mathcal{K}, \label{Cachenable09i} \\
&V(\phi_k)\leq\psi_k, \forall k\in\mathcal{K}, \label{Cachenable09j} \\
&\sqrt{\psi_k}\leq\theta_k, \forall k\in\mathcal{K}, \label{Cachenable09k}
\end{align}
\end{subequations}
\begin{figure*}[t]
\renewcommand{\captionfont}{\footnotesize}
\renewcommand*\captionlabeldelim{.}
	\centering
	\captionstyle{flushleft}
	\onelinecaptionstrue
	\includegraphics[width=1.0\columnwidth,keepaspectratio]{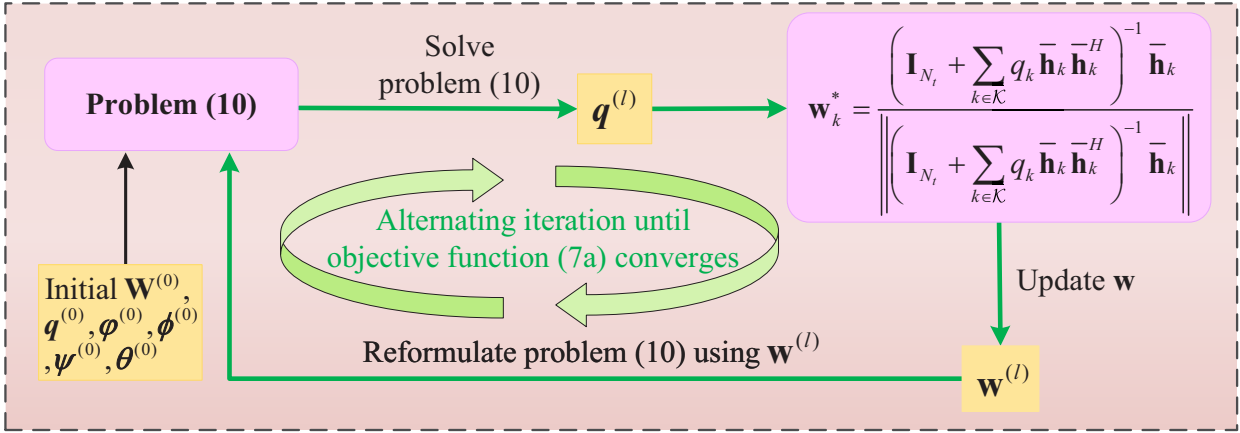}
	\caption{Illustration of alternating iteration algorithm proposed in~\cite{he2021beamforming}.}
	\label{HeBF}
\end{figure*}
In problem~\eqref{Cachenable09}, the optimization variables are $q_k, \varphi_k, \phi_k, \psi_k$, and $\theta_k, \forall k\in\mathcal{K}$. $\tilde{\gamma}_k$ is given by $\tilde{\gamma}_k=\frac{P\|\mathbf{h}_k^{H}\|^2}{\sigma_k^2}$. $\nu_3$ is the solution of $\overleftarrow{R}\left(\overleftarrow{\gamma}\right)=\frac{D}{n}\ln2$, which is given in~\cite{he2021beamforming}. An alternating iteration algorithm based on SCA is proposed in~\cite{he2021beamforming}. Specifically, for given beamforming vector $\{\mathbf{w}_k\}$, the algorithm solves problem~\eqref{Cachenable09} to obtain $\{q_k\}$. Then, the beamforming vector $\{\mathbf{w}_k\}$ is updated using $\{q_k\}$ via~\eqref{Cachenable08}. The aforementioned solving procedure is implemented iteratively until the objective function~\eqref{Cachenable07a} converges, which is illustrated in Fig.~\ref{HeBF}. Although the proposed algorithm achieves good performance, but the computational complexity of it is relatively high, and there is still a certain distance from real-time applications. In order to propose a solution with lower computational complexity, we will design a learning model to solve problem~\eqref{Cachenable07} in the following subsection.

\subsection*{B. Unrolled SRMax Network~(USRMNet)}
\begin{figure*}[t]
\renewcommand{\captionfont}{\footnotesize}
\renewcommand*\captionlabeldelim{.}
	\centering
	\captionstyle{flushleft}
	\onelinecaptionstrue
	\includegraphics[width=1.0\columnwidth,keepaspectratio]{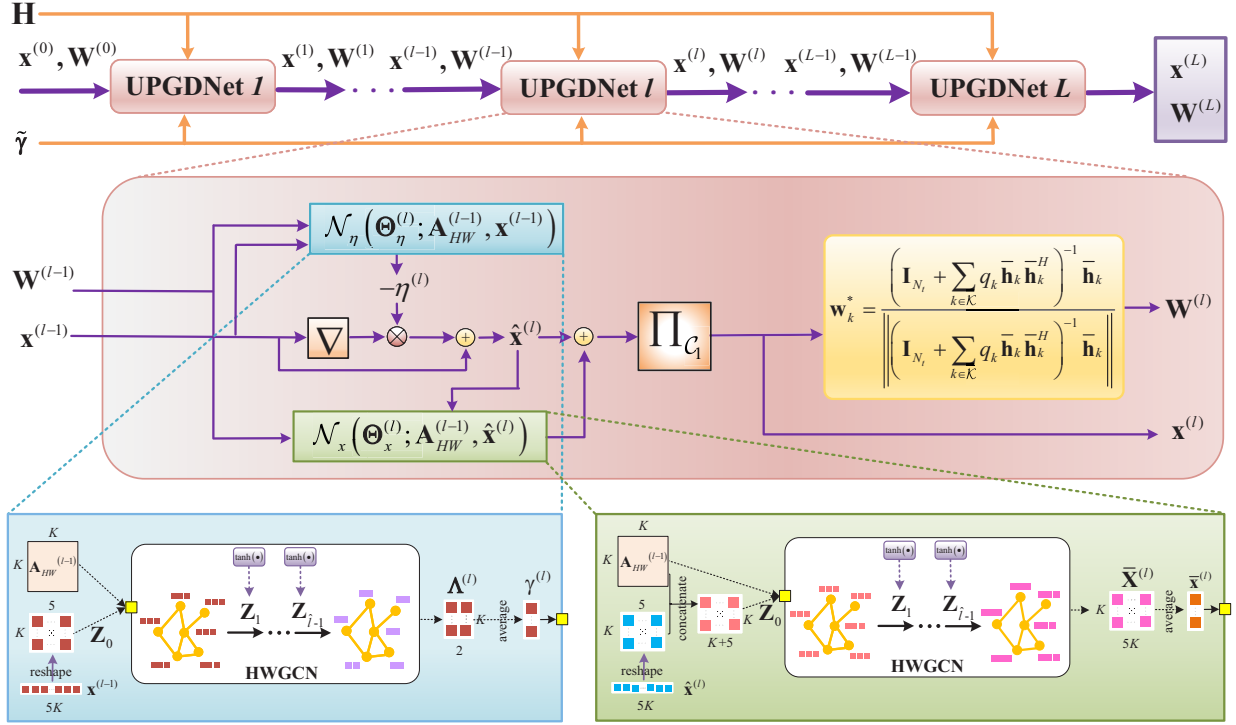}
	\caption{Illustration of the USRMNet model.}
	\label{USRMaxNet}
\end{figure*}

In this subsection, we focus on designing a learning model based on the UPGDNet and the analytical beamforming form to solve problem~\eqref{Cachenable07} via unrolling manner. Specifically, we unroll the alternating iteration algorithm proposed in~\cite{he2021beamforming} layer-by-layer. Each unrolled layer solves problem~\eqref{Cachenable09}, followed with an analytical beamforming update model with~\eqref{Cachenable08} at the end of the unrolled layer. Here, the unrolled layer acts as a UPGDNet. As a consequence, the unrolling learning model, i.e., USRMNet, is designed by stacking $L>0$ UPGDNets, which is illustrated in Fig.~\ref{USRMaxNet}. For the convenience of implementation, we concatenate the optimization variables in problem~\eqref{Cachenable09} into a vector $\mathbf{x}\in\mathbb{R}^{M}$, i.e., $\mathbf{x}=[\bm{q};\bm{\varphi};\bm{\phi};\bm{\psi};\bm{\theta}]^T$, where $M=5K, \bm{q}=[q_1,\dots, q_K]^T$, $\bm{\varphi}=[\varphi_1,\dots, \varphi_K]^T$, $\bm{\phi}=[\phi_1,\dots, \phi_K]^T$, $\bm{\psi}=[\psi_1,\dots, \psi_K]^T$, and $\bm{\theta}=[\theta_1,\dots, \theta_K]^T$. We also concatenate $\{\mathbf{w}_k\}$ into a matrix $\mathbf{W}=\left[\mathbf{w}_1^T;\cdots;\mathbf{w}_K^T\right]\in\mathbb{C}^{K\times N_{\mathrm{t}}}$. Further, for problem~\eqref{Cachenable09}, the constraint sets $\mathcal{C}_1$ and $\mathcal{C}_2$ are set as $\{\eqref{Cachenable09b}-\eqref{Cachenable09g}\}$ and $\{\eqref{Cachenable09h}-\eqref{Cachenable09k}\}$, respectively.
\begin{figure*}[t]
\renewcommand{\captionfont}{\footnotesize}
\renewcommand*\captionlabeldelim{.}
	\centering
	\captionstyle{flushleft}
	\onelinecaptionstrue
	\includegraphics[width=1.0\columnwidth,keepaspectratio]{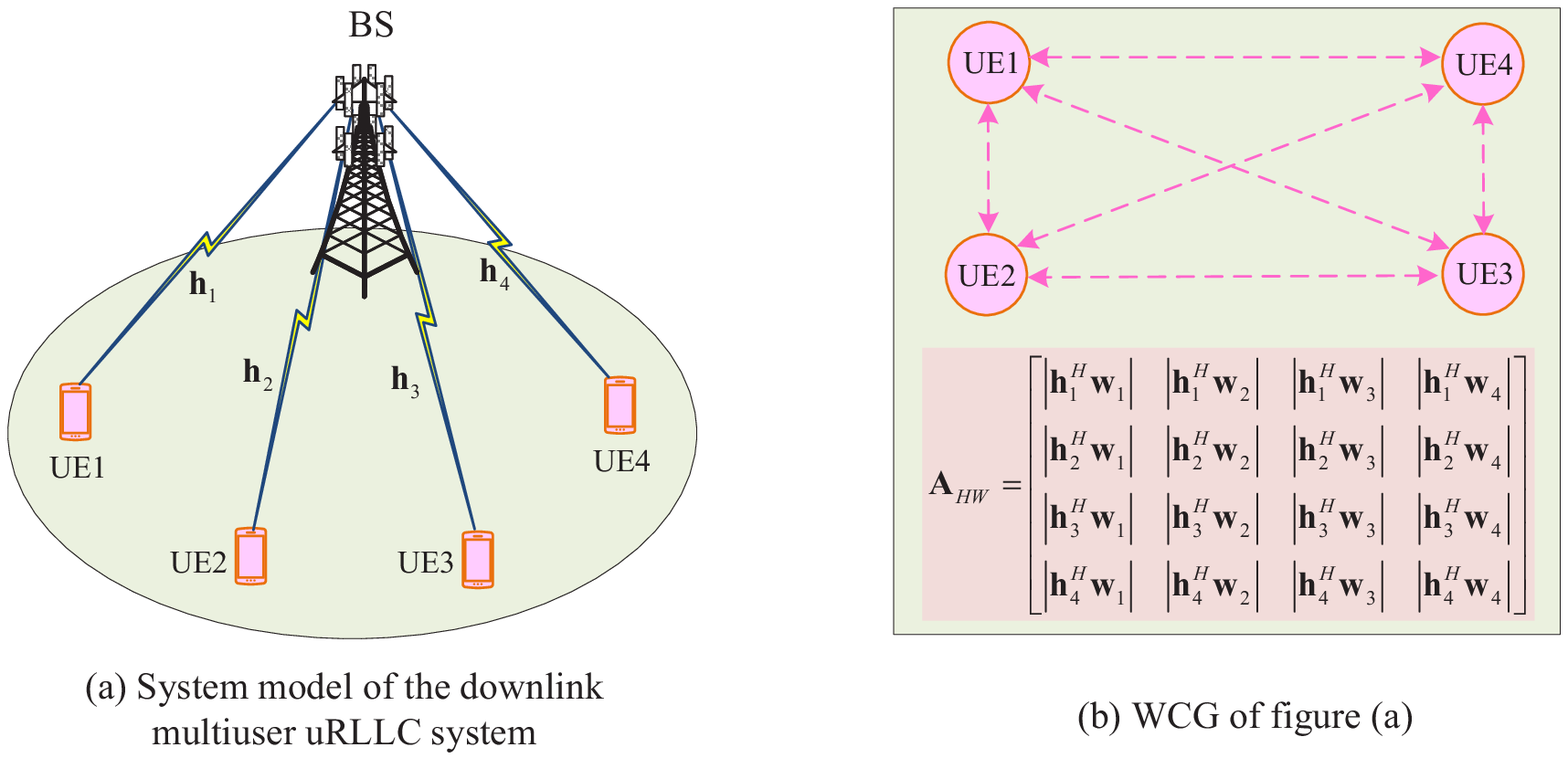}
	\caption{Illustration of WCG for HWGCN in the downlink multiuser uRLLC system.}
	\label{WCG}
\end{figure*}
In the USRMNet model, the initial inputs are $\mathbf{x}^{(0)}$ and $\mathbf{W}^{(0)}$, which are initialized via solving problem~(36) in~\cite{he2021beamforming}. The inputs of the $l$-th UPGDNet layer are the outputs of the $(l-1)$-th UPGDNet layer, i.e., $\mathbf{x}^{(l-1)},\mathbf{W}^{(l-1)}$. In the $l$-th UPGDNet layer, $\mathcal{N}_{\eta}\left(\mathbf{\Theta}_{\eta}^{(l)};\mathbf{A}_{HW}^{(l-1)},\mathbf{x}^{(l-1)}\right)$ and $\mathcal{N}_{x}\left(\mathbf{\Theta}_{x}^{(l)};\mathbf{A}_{HW}^{(l-1)}, \hat{\mathbf{x}}^{(l)}\right)$ are designed based on spectral-based graph convolutional NN, called HWGCNs, to generate gradient descent step-size vector $\bm{\eta}^{(l)}=\left[\eta_1^{(l)},\eta_2^{(l)}\right]^T\in\mathbb{R}^{2}$ and perturbation vector $\bar{\mathbf{x}}^{(l)}\in\mathbb{R}^{M}$, respectively~\footnote{Due to the preliminary estimation stage, i.e., step 3 in \textbf{Algorithm} 1, is only related to two optimization variables, namely $\bm{\varphi}$ and $\bm{\theta}$, this paper only designs two gradient descent step sizes $\eta_1$ and $\eta_2$ for these two optimization variables, respectively.}. Specifically, the HWGCNs utilized in $\mathcal{N}_{\eta}\left(\mathbf{\Theta}_{\eta}^{(l)};\mathbf{A}_{HW}^{(l-1)},\mathbf{x}^{(l-1)}\right)$ and $\mathcal{N}_{x}\left(\mathbf{\Theta}_{x}^{(l)};\mathbf{A}_{HW}^{(l-1)}, \hat{\mathbf{x}}^{(l)}\right)$ output gradient descent step size matrix  $\mathbf{\Lambda}^{(l)}=\left[\bm{\eta}^{(l,1)},\dots,\bm{\eta}^{(l,K)}\right]^T\in\mathbb{R}^{K\times 2}$ and perturbation matrix $\overline{\mathbf{X}}^{(l)}=\left[\bar{\mathbf{x}}^{(l,1)},\dots,\bar{\mathbf{x}}^{(l,K)}\right]^T\in\mathbb{R}^{K\times 5K}$ for all UEs, respectively. $\bm{\eta}^{(l,k)}\in\mathbb{R}^{2}$ and $\bar{\mathbf{x}}^{(l,k)}\in\mathbb{R}^{M}$ are the gradient descent step-size vector and the perturbation vector for the $k$-th UE in the $l$-th UPGDNet layer, respectively. Then, $\bm{\eta}^{(l)}$ and $\bar{\mathbf{x}}^{(l)}$ can be obtained by $\bm{\eta}^{(l)}=\frac{1}{K}\sum_{k=1}^{K}\bm{\eta}^{(l,k)}$ and $\bar{\mathbf{x}}^{(l)}=\frac{1}{K}\sum_{k=1}^{K}\bar{\mathbf{x}}^{(l,k)}$, respectively.

In what follows, we focus on illustrating the design of HWGCN. We begin with building a wireless communication graph~(WCG) by regarding each UE as a vertex corresponding to the downlink multiuser uRLLC system, which is illustrated in Fig.~\ref{WCG}. The adjacent matrix of WCG in the $l$-the UPGDNet layer of the USRMNet model is defined as $\mathbf{A}_{HW}^{(l)}=|\mathbf{H}\left(\mathbf{W}^{(l)}\right)^T|$, where $\mathbf{H}=\left[\mathbf{h}_1,\cdots,\mathbf{h}_K\right]^H\in\mathbb{C}^{K\times N_{\mathrm{t}}}$. The intermediate feature vector $\mathbf{Z}_{\hat{l}}\in\mathbb{R}^{K\times p}$ in the $\hat{l}$-th layer of HWGCN is generated as follows
\begin{equation}\label{Cachenable10}
\begin{aligned}
\mathbf{Z}_{\hat{l}}=\kappa_{\hat{l}}\left(\sum_{f=1}^{F_{\hat{l}}}\sum_{k=0}^{K_{\hat{l}}-1}\beta_{\hat{l}k}^{f}\left(\mathbf{A}_{HW}^{(l)}\right)^k\mathbf{Z}_{\hat{l}-1}^f\right),
\end{aligned}
\end{equation}
where $p$ is the dimension of immediate feature vector of each vertex, $\beta_{\hat{l}k}^f$ denotes the $k$-th filter coefficient of the $f$-th graph filter bank in the $\hat{l}$-th layer. $F_{\hat{l}}$ and $K_{\hat{l}}$ denote the number of the graph filter banks and the number of filter coefficients of each graph filter bank in the $\hat{l}$-th layer, respectively. $\kappa_{\hat{l}}$ denotes the nonlinear activation function in the $\hat{l}$-th layer of HWGCN. The inputs of $\mathcal{N}_{\eta}\left(\mathbf{\Theta}_{\eta}^{(l)};\mathbf{A}_{HW}^{(l-1)},\mathbf{x}^{(l-1)}\right)$ and $\mathcal{N}_{x}\left(\mathbf{\Theta}_{x}^{(l)};\mathbf{A}_{HW}^{(l-1)}, \hat{\mathbf{x}}^{(l)}\right)$ are designed as $\Gamma\left(\mathbf{x}^{(l-1)}\right)\in\mathbb{R}^{K\times 5}$ and $[\mathbf{A}_{HW}^{(l-1)}, \Gamma\left(\hat{\mathbf{x}}^{(l)}\right)]\in\mathbb{R}^{K\times (K+5)}$, respectively. $\Gamma\left(\mathbf{x}\right):~\mathbb{R}^{5K}\rightarrow\mathbb{R}^{K\times 5}$ denotes the dimension transformation operation for $\mathbf{x}$, i.e., $\mathbf{x}=[\bm{q};\bm{\varphi};\bm{\phi};\bm{\psi};\bm{\theta}]^T\rightarrow
[\bm{q},\bm{\varphi},\bm{\phi},\bm{\psi},\bm{\theta}]$.
\begin{remark}
Generally speaking, $\mathcal{N}_{\eta}\left(\mathbf{\Theta}_{\eta}^{(l)};\mathbf{A}_{HW}^{(l-1)},\mathbf{x}^{(l-1)}\right)$ and $\mathcal{N}_{x}\left(\mathbf{\Theta}_{x}^{(l)};\mathbf{A}_{HW}^{(l-1)}, \hat{\mathbf{x}}^{(l)}\right)$ can be designed as various DNNs, which are not limited to GNNs. The reasons for choosing GNN in this paper are as follows. On one hand, GNN has the property of PE, which meet the disorder of UEs in wireless networks. On the other hand, GNN acts on the graph domain and is suitable for the natural topology of wireless networks, which is appropriate to exploit non-Euclidean data in wireless networks. Finally, the HWGCN utilized in this paper has a small number of learning parameters, which is helpful to reduce the training and testing complexities of the USRMNet model.
\end{remark}

\subsection*{C. PE of the USRMNet model}
In general, in the context of wireless networks, the UEs in the network are disordered. Reordering the UEs may influence the optimization outputs of the learning model constructed with general DNNs, e.g., fully connected NNs. Fortunately, in GNNs, the PE suggests that the permutation of vertices is independent of the output, i.e., the permutation of inputs leads to the same permutation of outputs~\cite{zhang2021scalable}. From the characteristics of wireless networks and the design of HWGCN, it is necessary to prove whether the USRMNet model has the PE property.

A permutation matrix of $K$ dimension is defined as $\mathbf{\Xi}=\{0, 1\}^{K\times K}$ with $\mathbf{\Xi}\mathbf{1}=\mathbf{1},\mathbf{\Xi}^T\mathbf{1}=\mathbf{1}$. The multiplication of a vector by permutation matrix, i.e., $\mathbf{\Xi}^T\mathbf{v}$ reorders the entries of vector $\mathbf{v}$. The multiplication $\mathbf{\Xi}^T\mathbf{Q}\mathbf{\Xi}$ reorders the rows and columns of any given matrix $\mathbf{Q}$. A utility function $\pi:\mathbb{R}^{K\times K}\rightarrow\mathbb{R}^{K}$ is permutation equivariant if $\pi\left(\mathbf{\Xi}^T\mathbf{Q}\mathbf{\Xi}\right)=\mathbf{\Xi}^T\pi\left(\mathbf{Q}\right)$ for all matrices $\mathbf{Q}$ and all permutation matrices $\mathbf{\Xi}$. For the convenience of description, we denote the USRMNet model as $\Upsilon\left(\mathbf{\Theta}_x, \mathbf{\Theta}_{\eta};\mathbf{A}_{HW}, \mathbf{W}, \mathbf{x}\right)$. While $\Upsilon^{(l)}\left(\mathbf{\Theta}_x^{(l)}, \mathbf{\Theta}_{\eta}^{(l)};\mathbf{A}_{HW}^{(l-1)}, \mathbf{W}^{(l-1)}, \mathbf{x}^{(l-1)}\right)$ indicates the $l$-th UPGDNet layer. Note that the input vector $\mathbf{x}$ is consist with optimization variables $\bm{q}, \bm{\varphi}, \bm{\phi}, \bm{\psi}$, and $\bm{\theta}$, the permutation operation is implemented on these optimization variables instead of $\mathbf{x}$. Then, the arbitrary permutations of $\mathbf{A}_{HW}$ and the inputs to the $l$-th UPGDNet layer are denoted by $\mathbf{\Xi}^T\mathbf{A}_{HW}\mathbf{\Xi}$, $\mathbf{\Xi}^T\bm{q}$, $\mathbf{\Xi}^T\bm{\varphi}$, $\mathbf{\Xi}^T\bm{\phi}$, $\mathbf{\Xi}^T\bm{\psi}$, $\mathbf{\Xi}^T\bm{\theta}$, and $\mathbf{\Xi}^T\mathbf{W}$, respectively. For ease of description, let $\mathbf{\Xi}^T\star\mathbf{x}$ denotes the multiplication of $\mathbf{x}$ by $\mathbf{\Xi}^T$, i.e., $\mathbf{\Xi}^T\star\mathbf{x}=[\mathbf{\Xi}^T\bm{q};\mathbf{\Xi}^T\bm{\varphi};\mathbf{\Xi}^T\bm{\phi};\mathbf{\Xi}^T\bm{\psi};\mathbf{\Xi}^T\bm{\theta}]\in\mathbb{R}^{5K}$.  In what follows, we discuss the PE of our proposed USRMNet model.

\begin{proposition}\label{proposition1}
  Given matrix $\mathbf{A}_{HW}$ and input $\mathbf{x}$, let $\mathbf{A}_{HW}^{'}=\mathbf{\Xi}^T\mathbf{A}_{HW}\mathbf{\Xi}$ and $\mathbf{x}^{'}=\mathbf{\Xi}^T\star\mathbf{x}$ for a permutation matrix $\mathbf{\Xi}$, $\mathcal{N}_{\eta}\left(\mathbf{\Theta}_{\eta};\mathbf{A}_{HW},\mathbf{x}\right)$ and $\mathcal{N}_{x}\left(\mathbf{\Theta}_{x};\mathbf{A}_{HW}, \hat{\mathbf{x}}\right)$ designed with HWGCN are permutation equivariant, i.e.,
  \begin{subequations}\label{Cachenable15}
    \begin{align}
    &\mathcal{N}_{\eta}\left(\mathbf{\Theta}_{\eta};\mathbf{A}_{HW}^{'},\mathbf{x}^{'}\right)=\mathbf{\Xi}^{T}\mathcal{N}_{\eta}\left(\mathbf{\Theta}_{\eta};\mathbf{A}_{HW},\mathbf{x}\right)=\mathbf{\Xi}^T\mathbf{\Lambda},~\label{Cachenable15a} \\
    &\mathcal{N}_{x}\left(\mathbf{\Theta}_{x};\mathbf{A}_{HW}^{'},\mathbf{x}^{'}\right)=\mathbf{\Xi}^{T}\mathcal{N}_{x}\left(\mathbf{\Theta}_{x};\mathbf{A}_{HW},\mathbf{x}\right)=\mathbf{\Xi}^T\overline{\mathbf{X}}.~\label{Cachenable15b}
    \end{align}
  \end{subequations}
\end{proposition}

\begin{IEEEproof}
Note that there is a common dimension transformation operation $\Gamma\left(\cdot\right)$ in $\mathcal{N}_{\eta}\left(\mathbf{\Theta}_{\eta};\mathbf{A}_{HW},\mathbf{x}\right)$ and $\mathcal{N}_{x}\left(\mathbf{\Theta}_{x};\mathbf{A}_{HW}, \hat{\mathbf{x}}\right)$. We should first guarantee the PE of $\Gamma\left(\cdot\right)$. From the definition of $\Gamma\left(\cdot\right)$, it is easy to have $\Gamma\left(\mathbf{\Xi}^T\star\mathbf{x}\right)=\mathbf{\Xi}^T\Gamma\left(\mathbf{x}\right)$ and $\Gamma\left(\mathbf{\Xi}^T\star\hat{\mathbf{x}}\right)=\mathbf{\Xi}^T\Gamma\left(\hat{\mathbf{x}}\right)$ as we wanted. Since the HWGCN utilized in $\mathcal{N}_{\eta}\left(\mathbf{\Theta}_{\eta};\mathbf{A}_{HW},\mathbf{x}\right)$ and $\mathcal{N}_{x}\left(\mathbf{\Theta}_{x};\mathbf{A}_{HW}, \hat{\mathbf{x}}\right)$ have similar structure, we only analyze the PE of $\mathcal{N}_{\eta}\left(\mathbf{\Theta}_{\eta};\mathbf{A}_{HW},\mathbf{x}\right)$ for convenience. For simplicity, considering the first layer $\hat{l}=1$, we assume the number of the graph filter banks $F_{\hat{l}}=1$ at each layer $\hat{l}$ of HWGCN in $\mathcal{N}_{\eta}\left(\mathbf{\Theta}_{\eta};\mathbf{A}_{HW},\mathbf{x}\right)$. According to the definition of HWGCN, i.e., equation~\eqref{Cachenable09}, take $\mathbf{A}_{HW}^{'}$ and $\Gamma\left(\mathbf{x}{'}\right)$ as inputs, we have
\begin{equation}\label{Cachenable16}
\begin{aligned}
\mathbf{Z}_{2}^{'}&=\kappa_{1}\left(\sum_{k=0}^{K_{1}-1}\beta_{1k}^{1}\left(\mathbf{A}_{HW}^{'}\right)^k\Gamma\left(\mathbf{x}^{'}\right)\right) =\kappa_{1}\left(\sum_{k=0}^{K_{1}-1}\beta_{1k}^{1}\left(\mathbf{\Xi}^T\mathbf{A}_{HW}\mathbf{\Xi}\right)^k\mathbf{\Xi}^T\Gamma\left(\mathbf{x}\right)\right).
\end{aligned}
\end{equation}
Since $\mathbf{\Xi}^T\mathbf{\Xi}=\mathbf{\Xi}\mathbf{\Xi}^T=\mathbf{1}$, we have $\left(\mathbf{\Xi}^T\mathbf{A}_{HW}\mathbf{\Xi}\right)^k=\mathbf{\Xi}^T\mathbf{A}_{HW}^k\mathbf{\Xi}$. Then, we reformulate~\eqref{Cachenable16} as ~\eqref{Cachenable17}.
\begin{equation}\label{Cachenable17}
\begin{aligned}
\mathbf{Z}_{2}^{'}&=\kappa_{1}\left(\sum_{k=0}^{K_{1}-1}\beta_{1k}^{1}\mathbf{\Xi}^T\mathbf{A}_{HW}^k\Gamma\left(\mathbf{x}\right)\right)=\kappa_{1}\left(\mathbf{\Xi}^T\sum_{k=0}^{K_{1}-1}\beta_{1k}^{1}\mathbf{A}_{HW}^k\Gamma\left(\mathbf{x}\right)\right).
\end{aligned}
\end{equation}
Due to $\kappa_{1}\left(\cdot\right)$ is a point-wise activation function, we have $\mathbf{Z}_{2}^{'}=\mathbf{\Xi}^T\kappa_{1}\left(\sum_{k=0}^{K_{1}-1}\beta_{1k}^{1}\mathbf{A}_{HW}^k\Gamma\left(\mathbf{x}\right)\right)=\mathbf{\Xi}^{T}\mathbf{Z}_{2}$, where $\mathbf{Z}_{2}$ is the output of the first layer of HWGCN under the inputs that are not permuted. The outputs of the current layer are permutation equivariant to the inputs of the next layer, so the other layer of HWGCN are also permutation equivariant, i.e., $\mathcal{N}_{\eta}\left(\mathbf{\Theta}_{\eta};\mathbf{A}_{HW}^{'},\mathbf{x}^{'}\right)=\mathbf{\Xi}^{T}\mathcal{N}_{\eta}\left(\mathbf{\Theta}_{\eta};\mathbf{A}_{HW},\mathbf{x}\right)=\mathbf{\Xi}^T\mathbf{\Lambda}$. Similarly, we have $\mathcal{N}_{x}\left(\mathbf{\Theta}_{x};\mathbf{A}_{HW}^{'},\mathbf{x}^{'}\right)=\mathbf{\Xi}^{T}\mathcal{N}_{x}\left(\mathbf{\Theta}_{x};\mathbf{A}_{HW},\mathbf{x}\right)=\mathbf{\Xi}^T\overline{\mathbf{X}}$.
\end{IEEEproof}

\begin{proposition}\label{proposition2}
  Given matrix $\mathbf{A}_{HW}$, inputs $\mathbf{x}$ and $\mathbf{W}$, for a permutation matrix $\mathbf{\Xi}$, let $\mathbf{A}_{HW}^{'}=\mathbf{\Xi}^T\mathbf{A}_{HW}\mathbf{\Xi}$, $\mathbf{x}^{'}=\mathbf{\Xi}^{T}\star\mathbf{x}$, and $\mathbf{W}^{'}=\mathbf{\Xi}^T\mathbf{W}$, $\Upsilon\left(\mathbf{\Theta}_x, \mathbf{\Theta}_{\eta};\mathbf{A}_{HW}, \mathbf{W}, \mathbf{x}\right)$ is also permutation equivariant, i.e.,
  \begin{equation}\label{Cachenable13}
    \begin{aligned}
    &\Upsilon\left(\mathbf{\Theta}_x, \mathbf{\Theta}_{\eta};\mathbf{A}_{HW}^{'}, \mathbf{W}^{'}, \mathbf{x}^{'}\right)=\mathbf{\Xi}^T\Upsilon\left(\mathbf{\Theta}_x, \mathbf{\Theta}_{\eta};\mathbf{A}_{HW}, \mathbf{W},\mathbf{x}\right).
    \end{aligned}
  \end{equation}
\end{proposition}

\begin{IEEEproof}
  First of all, we demonstrate the permutation $\mathbf{\Xi}^T\mathbf{A}_{HW}\mathbf{\Xi}$ is determined by the permutations $\mathbf{\Xi}^T\mathbf{H}$ and $\mathbf{\Xi}^T\mathbf{W}$. Since $\mathbf{\Xi}^T\mathbf{A}_{HW}\mathbf{\Xi}=\mathbf{\Xi}^T|\mathbf{H}\mathbf{W}^T|\mathbf{\Xi}=|\mathbf{\Xi}^T\mathbf{H}\mathbf{W}^T\mathbf{\Xi}|=|\mathbf{\Xi}^T\mathbf{H}\left(\mathbf{\Xi}^T\mathbf{W}\right)^T|$, $\mathbf{\Xi}^T\mathbf{A}_{HW}\mathbf{\Xi}$ corresponds to the permuted $\mathbf{\Xi}^T\mathbf{H}$ and $\mathbf{\Xi}^T\mathbf{W}$.

  Now, for simplicity, we are going to show that the $l$-th UPGDNet in the USRMNet model, i.e., $\Upsilon^{(l)}\left(\mathbf{\Theta}_x^{(l)}, \mathbf{\Theta}_{\eta}^{(l)};\mathbf{A}_{HW}^{(l-1)}, \mathbf{W}^{(l-1)}, \mathbf{x}^{(l-1)}\right)$ is permutation equivariant. Specifically, we should discuss \textbf{Algorithm} 1 and the beamforming vector update operation~\eqref{Cachenable08} whether have PE property. Let $\left\{\mathbf{x}^{(l)}, \mathbf{W}^{(l)}\right\}$ be the outputs of $\Upsilon^{(l)}\left(\mathbf{\Theta}_x^{(l)}, \mathbf{\Theta}_{\eta}^{(l)};\mathbf{A}_{HW}^{(l-1)}, \mathbf{W}^{(l-1)}, \mathbf{x}^{(l-1)}\right)$. For steps 2 and 5 in \textbf{Algorithm} 1, from the PE of $\mathcal{N}_{\eta}\left(\mathbf{\Theta}_{\eta};\mathbf{A}_{HW},\mathbf{x}\right)$ and $\mathcal{N}_{x}\left(\mathbf{\Theta}_{x};\mathbf{A}_{HW}, \hat{\mathbf{x}}\right)$, we have that $\mathcal{N}_{\eta}\left(\mathbf{\Theta}_{\eta}^{(l)};\mathbf{\Xi}^T\mathbf{A}_{HW}^{(l-1)}\mathbf{\Xi},\mathbf{\Xi}^T\star\mathbf{x}^{(l-1)}\right)=\mathbf{\Xi}^T\bm{\Lambda}^{(l)}$ and $\mathcal{N}_{x}^{(l)}\left(\mathbf{\Theta}_{x}^{(l)};\mathbf{\Xi}^T\mathbf{A}_{HW}^{(l-1)}\mathbf{\Xi},\mathbf{\Xi}^T\star\mathbf{x}^{(l)}\right)=\mathbf{\Xi}^T\overline{\mathbf{X}}^{(l)}$, where $\mathbf{\Lambda}^{(l)}$ and $\overline{\mathbf{X}}^{(l)}$ are the outputs that generated by $\mathbf{A}_{HW}$ and $\mathbf{x}$. For the step 3 in \textbf{Algorithm} 1, $\hat{\mathbf{x}}^{'} = \mathbf{\Xi}^T\star\mathbf{x}_0 - \eta\nabla_{x}f\left(\mathbf{\Xi}^T\star\mathbf{x}_0\right)=\mathbf{\Xi}\star\hat{\mathbf{x}}$. Similarly, the steps 6-7 in \textbf{Algorithm} 1 also have PE property. Therefore, it follows that the \textbf{Algorithm} 1 in $\Upsilon^{(l)}\left(\mathbf{\Theta}_x^{(l)}, \mathbf{\Theta}_{\eta}^{(l)};\mathbf{A}_{HW}^{(l-1)}, \mathbf{W}^{(l-1)}, \mathbf{x}^{(l-1)}\right)$ is permutation equivariant. While for the beamforming vector update operation~\eqref{Cachenable08}, given the output $\mathbf{\Xi}^T\mathbf{x}^{(l)}$, take $\mathbf{\Xi}^T\bm{q}^{(l)}$ as input, we have
  \begin{equation}\label{Cachenable19}
  \begin{aligned}
  &\frac{\left(\mathbf{I}_{N_{\mathrm{t}}}+\sum\limits_{k\in\mathcal{K}}\left[\mathbf{\Xi}^T\bm{q}\right]_k\left[\mathbf{\Xi}^T\overline{\mathbf{H}}\right]_k\left[\mathbf{\Xi}^T\overline{\mathbf{H}}\right]_k^H\right)^{-1}\left[\mathbf{\Xi}^T\overline{\mathbf{H}}\right]_k}{\left|\left|\left(\mathbf{I}_{N_{\mathrm{t}}}+\sum\limits_{k\in\mathcal{K}}\left[\mathbf{\Xi}^T\bm{q}\right]_k\left[\mathbf{\Xi}^T\overline{\mathbf{H}}\right]_k\left[\mathbf{\Xi}^T\overline{\mathbf{H}}\right]_k^H\right)^{-1}\left[\mathbf{\Xi}^T\overline{\mathbf{H}}\right]_k\right|\right|}=\left[\mathbf{\Xi}^T\mathbf{W}^{(l)}\right]_k^T,
  \end{aligned}
  \end{equation}
  where $\left[\cdot\right]_k$ denotes the operation that take the $k$-th row of a column vector or the $k$-th column of a matrix, and the matrix $\overline{\mathbf{H}}$ is defined as $\overline{\mathbf{H}}=\left[\overline{\mathbf{h}}_1,\cdots,\overline{\mathbf{h}}_K\right]^H\in\mathbb{C}^{K\times N_{\mathrm{t}}}$. Hence, the $l$-th UPGDNet layer is permutation equivariant, i.e.,
  \begin{equation}\label{Cachenable18}
  \begin{aligned}
  &\Upsilon^{(l)}\left(\mathbf{\Theta}_x^{(l)}, \mathbf{\Theta}_{\eta}^{(l)};\mathbf{\Xi}^T\mathbf{A}_{HW}^{(l-1)}\mathbf{\Xi}, \mathbf{\Xi}^T\mathbf{W}^{(l-1)}, \mathbf{\Xi}^T\star\mathbf{x}^{(l-1)}\right)\\
  &=\mathbf{\Xi}^T\Upsilon^{(l)}\left(\mathbf{\Theta}_x^{(l)}, \mathbf{\Theta}_{\eta}^{(l)};\mathbf{A}_{HW}^{(l-1)}, \mathbf{W}^{(l-1)}, \mathbf{x}^{(l-1)}\right)\\
  &=\left\{\mathbf{\Xi}^T\mathbf{x}^{(l)},\mathbf{\Xi}^T\mathbf{W}^{(l)}\right\}=\mathbf{\Xi}^T\left\{\mathbf{x}^{(l)},\mathbf{W}^{(l)}\right\}.
  \end{aligned}
  \end{equation}

  Then, we leverage this to demonstrate that $\Upsilon\left(\mathbf{\Theta}_x, \mathbf{\Theta}_{\eta};\mathbf{A}_{HW}, \mathbf{W}, \mathbf{x}\right)$ is permutation equivariant. Considering the case where $L=1$, take the permutations $\mathbf{A}_{HW}^{'}, \mathbf{W}^{'}$, and $\mathbf{x}^{'}$ as inputs, we have
  \begin{equation}\label{Cachenable14}
    \begin{aligned}
    &\Upsilon\left(\mathbf{\Theta}_x, \mathbf{\Theta}_{\eta};\mathbf{A}_{HW}^{'}, \mathbf{W}^{'}, \mathbf{x}^{'}\right)\\
    &=\Upsilon^{(1)}\left(\mathbf{\Theta}_x^{(1)}, \mathbf{\Theta}_{\eta}^{(1)};\mathbf{\Xi}^T\mathbf{A}_{HW}^{(0)}\mathbf{\Xi}, \mathbf{\Xi}^T\mathbf{W}^{(0)}, \mathbf{\Xi}^T\star\mathbf{x}^{(0)}\right)\\
    &=\left\{\mathbf{\Xi}^T\mathbf{x}^{(1)},\mathbf{\Xi}^T\mathbf{W}^{(1)}\right\}=\mathbf{\Xi}^T\left\{\mathbf{x}^{(l)},\mathbf{W}^{(l)}\right\}\\
    &=\mathbf{\Xi}^T\Upsilon^{(1)}\left(\mathbf{\Theta}_x^{(1)}, \mathbf{\Theta}_{\eta}^{(1)};\mathbf{A}_{HW}^{(0)}, \mathbf{W}^{(0)}, \mathbf{x}^{(0)}\right)\\
    &=\mathbf{\Xi}^T\Upsilon\left(\mathbf{\Theta}_x, \mathbf{\Theta}_{\eta};\mathbf{A}_{HW}, \mathbf{W}, \mathbf{x}^{'}\right).
    \end{aligned}
  \end{equation}
  Therefore, the USRMNet model with a single UPGDNet layer is permutation equivariant. The PE property for $L>1$ can be demonstrated via a simple induction argument here omitted.
\end{IEEEproof}

From the combination of propositions~\ref{proposition1} and~\ref{proposition2}, it follows that the USRMNet model is permutation equivariant. Of course, $\mathcal{N}_{\eta}\left(\mathbf{\Theta}_{\eta};\mathbf{A}_{HW},\mathbf{x}\right)$ and $\mathcal{N}_{x}\left(\mathbf{\Theta}_{x};\mathbf{A}_{HW}, \hat{\mathbf{x}}\right)$ are not limited to be designed as HWGCN, while they can be designed based on arbitrary NNs with PE property.

\subsection*{D. Training process of the USRMNet model}
In this subsection, we illustrate the scheme to train the USRMNet model. Specifically, we train the USRMNet layer-by-layer in an unsupervised manner. Specifically, each UPGDNet is trained in the way of \textbf{Algorithm}~\ref{alg:algorithm2} by minimizing the objective function
\begin{equation}\label{Cachenable12}
\begin{aligned}
\mathcal{L}&=e^{-\frac{1}{2}s_1}\mathbb{E}_{\bm{\theta}}\{e^{-\sum\limits_{k\in\mathcal{K}}\alpha_k(\mathrm{ln}(1+\varphi_k)-\vartheta\theta_k)}\} +e^{-\frac{1}{2}s_2}\mathbb{E}_{\bm{\theta}}\{\sum\limits_{k\in\mathcal{K}}\bm{\lambda}_k^{\eqref{Cachenable09h}}[\varphi_k-\overleftarrow{\gamma}_k]^++\sum\limits_{k\in\mathcal{K}}\bm{\lambda}_k^{\eqref{Cachenable09i}}[\overleftarrow{\gamma}_k-\phi_k]^+\\
&+\sum\limits_{k\in\mathcal{K}}\bm{\lambda}_k^{\eqref{Cachenable09j}}[V\left(\phi_k\right)-\psi_k]^++\sum\limits_{k\in\mathcal{K}}\bm{\lambda}_k^{\eqref{Cachenable09k}}[\sqrt{\psi_k}-\theta_k]^+\}+e^{\frac{1}{2}s_1}+e^{\frac{1}{2}s_2},
\end{aligned}
\end{equation}
where $\bm{\lambda}^{\eqref{Cachenable09h}}\in\mathbb{R}^K_+$, $\bm{\lambda}^{\eqref{Cachenable09i}}\in\mathbb{R}^K_+$, $\bm{\lambda}^{\eqref{Cachenable09j}}\in\mathbb{R}^K_+$ and $\bm{\lambda}^{\eqref{Cachenable09k}}\in\mathbb{R}^K_+$ are Lagrangian multiplier vectors corresponding to constraints~\eqref{Cachenable09h}-\eqref{Cachenable09k}. $s_1$ and $s_2$ are the scale parameters for objective function~\eqref{Cachenable09a} maximization task and Lagrangian multiplier loss item minimization task, respectively. After training each UPGDNet layer, the learnable parameters of the current layer will be frozen, and beamforming vector $\{\mathbf{w}_{k}\}$ should be updated using~\eqref{Cachenable09} without calculating gradient. Accordingly, the constraints in problem~\eqref{Cachenable09} also should be adjusted with the updated $\{\mathbf{w}_{k}\}$.

\begin{remark}
For the design of HWGCN in the USRMNet model, the coefficient matrix $\mathbf{A}_{HW}^{(l)}$ is changing as the UPGDNet layer changes. The coefficient matrix $\mathbf{A}_{HW}^{(l)}$ reflects the state of vertices themselves and between vertices to a certain extent. Therefore, the WCG is dynamic graph in the USRMNet model.
\end{remark}

\section*{\sc \uppercase\expandafter{\romannumeral4}. Numerical Results}
\begin{figure*}[t]
\renewcommand{\captionfont}{\footnotesize}
\renewcommand*\captionlabeldelim{.}
	\centering
	\captionstyle{flushleft}
	\onelinecaptionstrue
	\includegraphics[width=1.0\columnwidth,keepaspectratio]{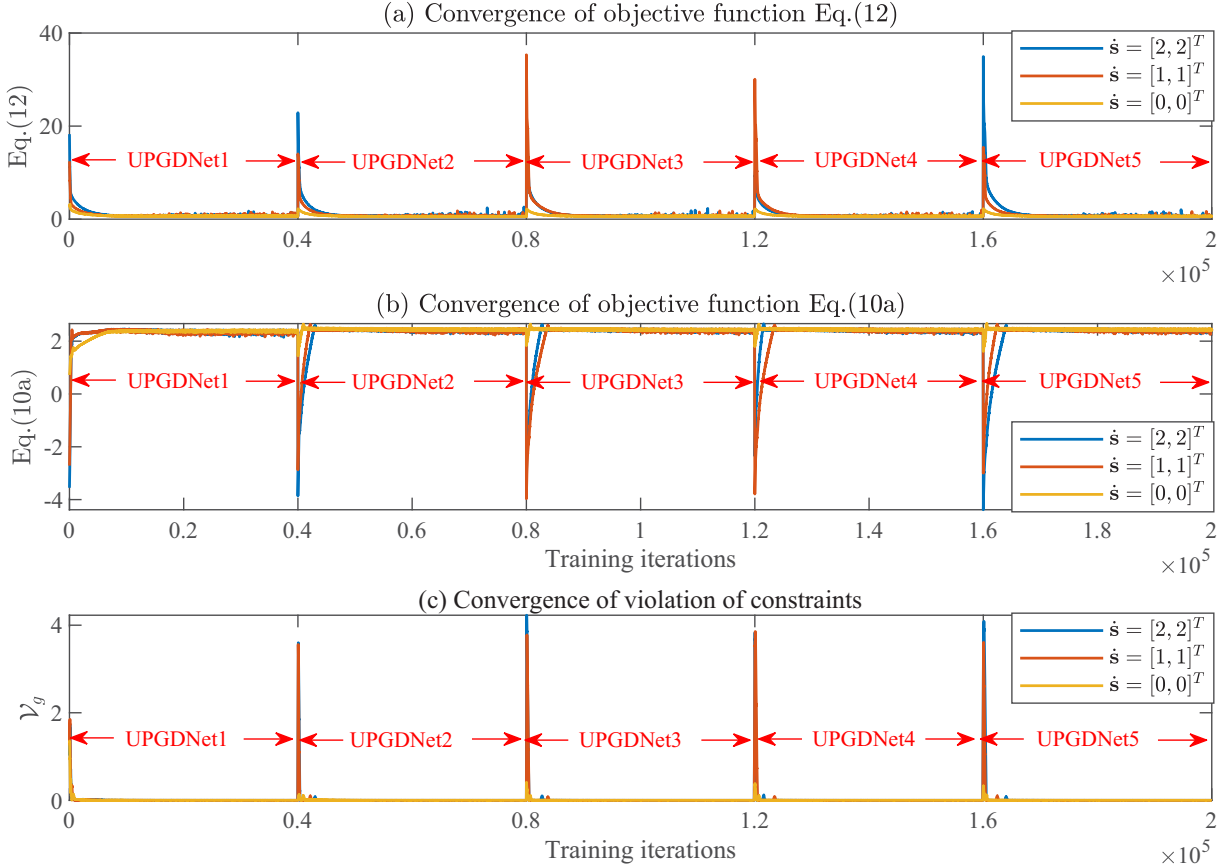}
	\caption{Illustration of convergence behaviors of objective function and violation of constraints during the training of the USRMNet model.}
	\label{convergence_obj}
\end{figure*}
\begin{figure*}[t]
\renewcommand{\captionfont}{\footnotesize}
\renewcommand*\captionlabeldelim{.}
	\centering
	\captionstyle{flushleft}
	\onelinecaptionstrue
	\includegraphics[width=1.0\columnwidth,keepaspectratio]{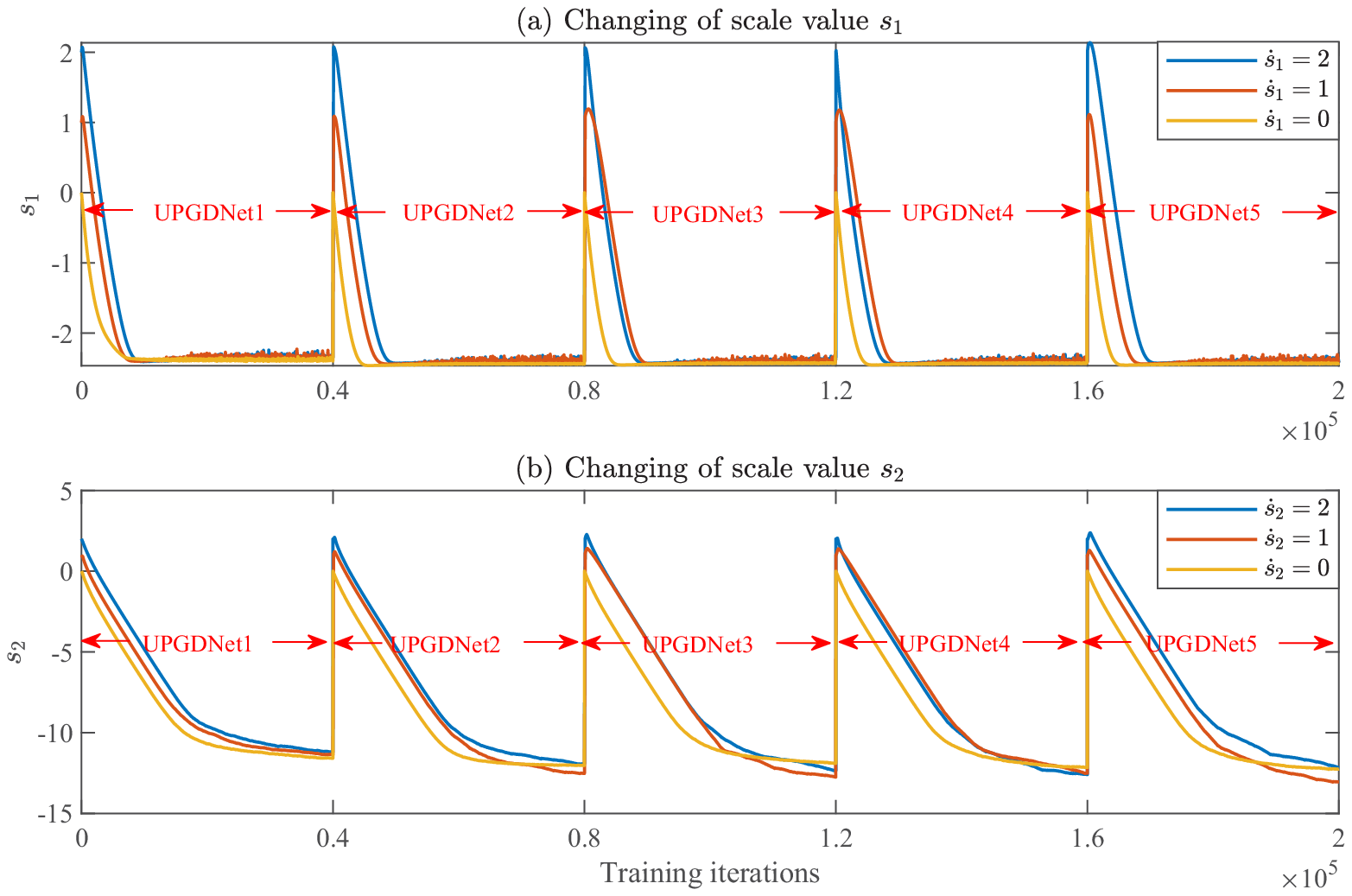}
	\caption{Illustration of convergence behaviors of scale values during the training of the USRMNet model.}
	\label{convergence_s}
\end{figure*}
In this section, to evaluate the effectiveness of the proposed USRMNet model for WSRMax problem in the downlink multiuser uRLLC system with finite blocklength transmission. The simulation model we considered consists of a single multi-antennas BS and $K$ single antenna UEs. The channel coefficient $\mathbf{h}_k$ from the BS to the $k$-th UE is modeled as $\mathbf{h}_k=\sqrt{\rho_k}\tilde{\mathbf{h}}_k$. $\rho_k$ denotes the channel power and it is defined as $\rho_k=1/\left(1+\left(d_k/d_0\right)^{\rho}\right)$, where $d_k$ is the distance between the BS and the $k$-th UE, $d_0$ and $\rho$ denote the reference distance and the fading exponent, respectively. The elements of $\tilde{\mathbf{h}}_k$ are independent and identically distributed (i.i.d) with $\mathcal{CN}\left(0,1\right)$. The radius of cell is denoted as $d_c$, and the minimum distance between the BS and UE is denoted as $d_u$. All UEs have the same noise variance, i.e., $\sigma_k^2=\sigma^2, \forall k\in\mathcal{K}$. For easy of notation, we define the SNR as $\text{SNR}=10\mathrm{log}_{10}\left(\frac{P}{\sigma^2}\right)$ in dB. 

In the following simulations, the $\mathcal{N}_{\eta}\left(\mathbf{\Theta}_{\eta};\mathbf{A}_{HW},\mathbf{x}\right)$ and $\mathcal{N}_{x}\left(\mathbf{\Theta}_{x};\mathbf{A}_{HW},\hat{\mathbf{x}}\right)$ of each UPGDNet are constructed with two HWGCNs, and the intermediate feature dimensions are designed as $\{5, 32, 2\}$ and $\{K+5, 32, 5\}$, respectively. The number of filter coefficients $K_{\hat{l}}=1$. The activation function utilized by the hidden layers is $\mathrm{tanh\left(x\right)}=\left(e^x-e^{-x}\right)/\left(e^x+e^{-x}\right)$. The final layer of $\mathcal{N}_{\eta}\left(\mathbf{\Theta}_{\eta};\mathbf{A}_{HW},\mathbf{x}\right)$ employs $\mathrm{ReLU}$ as activation function, while the final layer of $\mathcal{N}_{x}\left(\mathbf{\Theta}_{x};\mathbf{A}_{HW},\hat{\mathbf{x}}\right)$ does not use activation function. Each UPGDNet layer of the USRMNet model is trained for 50 epochs with $N_{train}=1\times10^4$ samples and tested with $N_{test}=5\times10^3$ samples. Specifically, we set $\varepsilon_{\lambda}=0.0001$, $\varepsilon_{s}=0.001$, $\varepsilon_{\Theta}=0.001$, and the mini-batch size $b=20$. The initial inputs are $\mathbf{x}^{(0)}$ and $\mathbf{W}^{(0)}$ are initialized via solving problem~(36) in~\cite{he2021beamforming}. The baseline scheme, i.e., the sub-optimal solution to problem~\eqref{Cachenable07} is obtained via Algorithm 1 that provided in~\cite{he2021beamforming}, labeled as HeBF, is considered for comparison.

\subsection*{A. Effectiveness of the USRMNet model}
\begin{figure*}[t]
\renewcommand{\captionfont}{\footnotesize}
\renewcommand*\captionlabeldelim{.}
	\centering
	\captionstyle{flushleft}
	\onelinecaptionstrue
	\includegraphics[width=0.8\columnwidth,keepaspectratio]{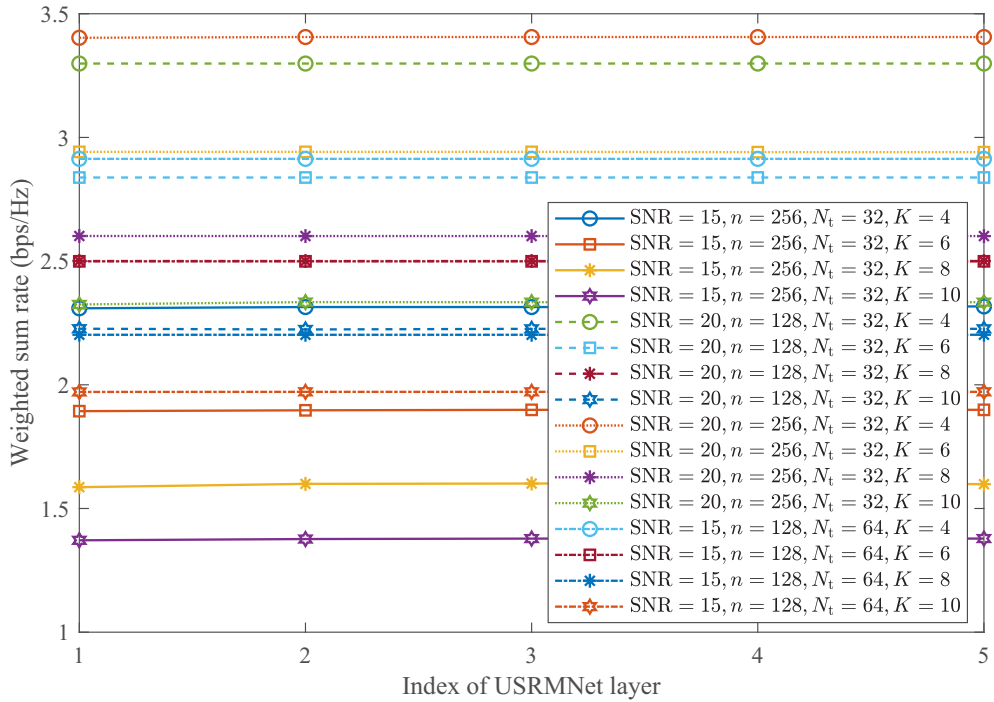}
	\caption{Illustration of convergence behaviors of the USRMNet model in terms of UPGDNet layer.}
	\label{convergence12}
\end{figure*}

In this subsection, we focus on evaluating the effectiveness of the learning framework of the UPGDNet and determining the layer number of the USRMNet model. Suppose the channel coefficients are generated with $\rho=3$, $d_0=50\mathrm{m}$, $d_u=120\mathrm{m}$, $d_c=140\mathrm{m}$, $K=6$, $\alpha_k=\frac{1}{K}$, $N_{\mathrm{t}}=64$, $\text{SNR}=15$ dB, $\varepsilon=10^{-5}$, $n=128$, and $D=256$ bits. To evaluate the influence of the initial value of $\mathbf{s}$, denoted as $\dot{\mathbf{s}}=[\dot{s}_1, \dot{s}_2]^T$, on the training of the USRMNet model, we train several USRMNet models with different $\dot{\mathbf{s}}$. Specifically, $\dot{\mathbf{s}}$ is set as $\left\{[2,2]^T,[1,1]^T,[0,0]^T\right\}$. The USRMNet model is constructed with 5 UPGDNets. As shown in Fig.~\ref{convergence_obj}, the objective functions~\eqref{Cachenable09a},~\eqref{Cachenable12}, and the violation of constraints can converge to stable point for the training of each UPGDNet in the USRMNet model with different $\dot{\mathbf{s}}$. Specifically, the violation of constraints is defined as $\mathcal{V}_g=\frac{\sum_{b=1}^{N_{test}}\mathcal{V}_b}{ N_{test}}$, where $\mathcal{V}_b=\frac{1}{4K}\{\sum\limits_{k\in\mathcal{K}}\left[\varphi_k^{(b)}-\overleftarrow{\gamma}_k^{(b)}\right]^+ +\sum\limits_{k\in\mathcal{K}}\left[\overleftarrow{\gamma}_k^{(b)}-\phi_k^{(b)}\right]^+ +\sum\limits_{k\in\mathcal{K}}\left[V\left(\phi_k^{(b)}\right)-\psi_k^{(b)}\right]^+ +\sum\limits_{k\in\mathcal{K}}\left[\sqrt{\psi_k^{(b)}}-\theta_k^{(b)}\right]^+\}$ is the violation of constraints of the $b$-th testing sample. In addition, we also analyzed the convergence behavior of scale parameters $\dot{\mathbf{s}}$ to demonstrate the effectiveness of the learning framework we proposed. As shown in Fig.~\ref{convergence_s}, with different $\dot{\mathbf{s}}$, $s_1$ and $s_2$ can converge to a similar stable point as the iteration number increases. Therefore, the proposed learning framework, i.e., Algorithm 2, and the multi-task objective function~\eqref{joint} are effective. To determine a proper number of UPGDNets for the USRMNet model, we also evaluate the performance of trained USRMNet model with test samples in terms of the objective function~\eqref{Cachenable07a}. These USRMNet model are trained with different simulation configurations, i.e., SNR is $\{15, 20\}$, $K$ is $\{4, 6, 8, 10\}$, $N_{\mathrm{t}}$ is $\{32, 64\}$, and $n$ is $\{128, 256\}$. Fig.~\ref{convergence12} shows the convergence behavior of the USRMNet model as the UPGDNet layer increases. It is not hard to find that the objective value basically does not increase when UPGDNet reaches layer 2. Among the $5\times10^3$ testing samples, $99\%$ of the samples have $\mathcal{V}_g=0$. We only consider the testing samples with $\mathcal{V}_g=0$ in the evaluation, and the following simulations are also the same. According to the experimental results, we would like to utilize the USRMNet model consisting of $2$ UPGDNets and set $\dot{\mathbf{s}}=[1,1]^T$ in the following numerical simulations.

\begin{table*} [htbp]
  \renewcommand{\tablename}
  \caption{\centering{ TABLE \uppercase\expandafter{\romannumeral1}} \protect \\ \quad\quad\quad\quad\quad\quad\quad\quad\quad WSR PERFORMANCE WITH DIFFERENT $K$, $n$, $N_{\mathrm{t}}$ and SNR}
  \\ \\
  \centering
  \begin{threeparttable}[b]
  \setlength{\tabcolsep}{4.0mm}{
    \begin{tabular}{| c | c | c | c | c | c | c | c |} \hline
 \multirow{2}{*}{\makecell[c]{SNR\\(dB)}} & \multirow{2}{*}{\makecell[c]{$N_{\mathrm{t}}$}} & \multirow{2}{*}{$n$} & & \multicolumn{4}{c|}{$K$} \\ \cline{5-8}
 & & & & 4 & 6 & 8 & 10  \\
 \hline
 \multirow{8}{*}{15} & \multirow{4}{*}{32} & \multirow{4}{*}{256} & \multirow{1}{*}{HeBF} & $2.3211$ & $1.9033$ & $1.6058$ & $1.3801$  \\
 \cline{4-8}
 & & & \multirow{1}{*}{USRMNet} & $2.3181$ & $1.8999$ & $1.6023$ & $1.3785$  \\
 \cline{4-8}
 & & & \multirow{1}{*}{$\varpi_1$} & $99.87\%$ & $99.82\%$ & $99.78\%$ & $99.88\%$  \\
 \cline{4-8}
 & & & \multirow{1}{*}{$\varpi_2$} & $100.00\%$ & $100.00\%$ & $100.00\%$ & $100.00\%$  \\
 \cline{2-8}
 & \multirow{4}{*}{64} & \multirow{4}{*}{128} & \multirow{1}{*}{HeBF} & $2.9138$ & $2.4997$ & $2.2029$ & $1.9726$  \\
 \cline{4-8}
 & & & \multirow{1}{*}{USRMNet} & $2.9134$ & $2.4993$ & $2.2023$ & $1.9717$  \\
 \cline{4-8}
 & & & \multirow{1}{*}{$\varpi_1$} & $99.99\%$ & $99.98\%$ & $99.97\%$ & $99.95\%$  \\
 \cline{4-8}
 & & & \multirow{1}{*}{$\varpi_2$} & $100.00\%$ & $100.00\%$ & $100.00\%$ & $100.00\%$  \\
 \hline
 \multirow{8}{*}{20} & \multirow{8}{*}{32} & \multirow{4}{*}{128} & \multirow{1}{*}{HeBF} & $3.3012$ & $2.8436$ & $2.5033$ & $2.2300$  \\
 \cline{4-8}
 & &  & \multirow{1}{*}{USRMNet} & $3.2987$ & $2.8381$ & $2.5005$ & $2.2267$  \\
 \cline{4-8}
 & &  & \multirow{1}{*}{$\varpi_1$} & $99.92\%$ & $99.81\%$ & $99.89\%$ & $99.85\%$  \\
 \cline{4-8}
 & & & \multirow{1}{*}{$\varpi_2$} & $100.00\%$ & $99.98\%$ & $100.00\%$ & $100.00\%$  \\
 \cline{3-8}
 & & \multirow{4}{*}{256} & \multirow{1}{*}{HeBF} & $3.4147$ & $2.9543$ & $2.6146$ & $2.3387$  \\
 \cline{4-8}
 & & & \multirow{1}{*}{USRMNet} & $3.4059$ & $2.9417$ & $2.6026$ & $2.3343$  \\
 \cline{4-8}
 & & & \multirow{1}{*}{$\varpi_1$} & $99.74\%$ & $99.57\%$ & $99.54\%$ & $99.81\%$  \\
 \cline{4-8}
 & & & \multirow{1}{*}{$\varpi_2$} & $100.00\%$ & $99.96\%$ & $99.98\%$ & $99.88\%$  \\
 \hline
    \end{tabular}}
 \end{threeparttable}
\end{table*}

\subsection*{B. Scalability of the USRMNet model}
In this subsection, to test the scalability of the USRMNet model, we would like to compare the USRMNet model with HeBF with varying $K$, $N_{\mathrm{t}}$, SNR, and $n$. Specifically, we train several USRMNet models with the same system configurations as Subsection V. A, while changing SNR as $\{15, 20\}$, $N_{\mathrm{t}}$ as $\{32, 64\}$, $K$ as $\{4, 6, 8, 10\}$, and $n$ as $\{128, 256\}$. Then, we test the performance of the trained USRMNet models with testing datasets. TABLE \uppercase\expandafter{\romannumeral1} shows the
WSRs achieved by the USRMNet model and HeBF. The USRMNet model is also compared to HeBF in terms of the objective function~\eqref{Cachenable07a}. We define two ratios as $\varpi_1=\frac{\text{WSR achieved by USRMNet}}{\text{WSR achieved by HeBF}}\%$ and $\varpi_2=\frac{\text{\# testing samples with $\mathcal{V}_g=0$}}{N_{test}}\%$. It is not hard to find from TABLE \uppercase\expandafter{\romannumeral1}, for the various simulation configurations, the WSR achieved by the USRMNet model can always reach more than $99\%$ of that achieved by HeBF. In addition, the ratio $\varpi_2$ can also always reach more than $99\%$ with all the considered simulation configurations, which indicates that the USRMNet model has a good generalization ability. Therefore, the USRMNet model achieves close performance to that of HeBF with good constraint satisfaction.

\subsection*{C. Generalize to Varying UE Distributions}
\begin{figure*}[t]
\renewcommand{\captionfont}{\footnotesize}
\renewcommand*\captionlabeldelim{.}
	\centering
	\captionstyle{flushleft}
	\onelinecaptionstrue \includegraphics[width=0.8\columnwidth,keepaspectratio]{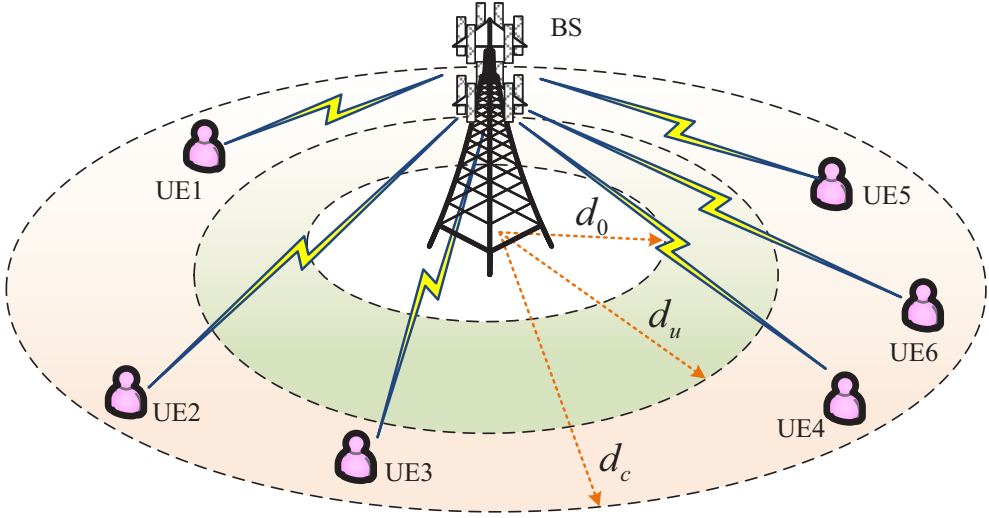}
	\caption{Illustration of UE distribution in the downlink multiuser uRLLC system.}
	\label{Cell}
\end{figure*}

In this subsection, we would like to test the generalization ability of the USRMNet model in terms of varying UE distributions. The so-called UE distribution refers to the maximum distance and minimum distance between UE and BS. Specifically, we train several USRMNet models with the same system configurations as Subsection V. A, while changing $K$ as $\{4, 6\}$ and $(d_u, d_c)$-train as $\{(120~\text{m}, 140~\text{m}), (100~\text{m}, 140~\text{m})\}$ and setting $N_{\mathrm{t}}=32, n=256$. Then, we utilize the pre-trained USRMNet models to test the samples with different UE distributions $\left(d_u, d_c\right)$-test from the pre-trained USRMNet models, while other system configurations are the same. Here, $(d_u, d_c)$-train and $(d_u,d_c)$-test denote the UE distribution used in the training samples for the pre-trained USRMNet model and the testing samples, respectively. The UE distribution in the downlink multiuser uRLLC system is shown in Fig.~\ref{Cell}. The USRMNet model is also compared to HeBF in terms of the objective function~\eqref{Cachenable07a}. For the scenario of $K=6$ with UE distribution $(100~\text{m}, 140~\text{m})$, to guarantee the satisfaction of constraints, we utilize the pre-trained USRMNet model with one UPGDNet layer. The experimental results are shown in TABLE \uppercase\expandafter{\romannumeral2}. We can observe that the WSR achieved by the pre-trained USRMNet model can reach more than $98\%$ of that achieved by HeBF in different UE distributions. For the violation of constraints $\mathcal{V}_g$, the ratio $\varpi_2$ can always reach more than $98\%$ with the training UE distributions $(120~\text{m}, 140~\text{m})$ and $(100~\text{m}, 140~\text{m})$. It shows that the USRMNet model could maintain a good generalization ability within a proper UE distribution different from that of the pre-trained USRMNet model in terms of varying UE numbers and UE distributions.
\begin{table*} [htbp]
  \renewcommand{\tablename}
  \caption{\centering{ TABLE \uppercase\expandafter{\romannumeral2}} \protect \\ \quad\quad\quad\quad\quad\quad\quad GENERALIZATION ABILITY ANALYSIS IN TERMS OF VARYING UE DISTRIBUTIONS}
  \\ \\
  \centering
  \begin{threeparttable}[b]
  \setlength{\tabcolsep}{3.0mm}{
    \begin{tabular}{| c | c | c | c | c | c | c |} \hline
  \multirow{1}{*}{$\left(d_u, d_c\right)$-train} & \multirow{1}{*}{$K$} & & \multicolumn{4}{c|}{$\left(d_u,d_c\right)$-test} \\ \cline{1-7}
  & & & (100~m, 120~m) & (140~m, 160~m) & (160~m, 180~m) & (180~m, 200~m)\\
 \cline{2-7}
 \multirow{8}{*}{(120, 140)} & \multirow{4}{*}{4} & \multirow{1}{*}{HeBF} & $2.7547$ & $1.9507$ & $1.6392$ & $1.3757$\\
 \cline{3-7}
 & & \multirow{1}{*}{USRMNet} & $2.7506$ & $1.9460$ & $1.6308$ & $1.3697$\\
 \cline{3-7}
 & & \multirow{1}{*}{$\varpi_1$} & $99.85\%$ & $99.76\%$ & $99.49\%$ & $99.56\%$\\
 \cline{3-7}
 & & \multirow{1}{*}{$\varpi_2$} & $100.00\%$ & $100.00\%$ & $99.60\%$ & $99.70\%$\\
 \cline{2-7}
  & \multirow{4}{*}{6} & \multirow{1}{*}{HeBF} & $2.3147$ & $1.5570$ & $1.2731$ & $1.0414$\\
 \cline{3-7}
 & & \multirow{1}{*}{USRMNet} & $2.3085$ & $1.5535$ & $1.2662$ & $1.0357$\\
 \cline{3-7}
 & & \multirow{1}{*}{$\varpi_1$} & $99.73\%$ & $99.78\%$ & $99.46\%$ & $99.45\%$\\
 \cline{3-7}
 & & \multirow{1}{*}{$\varpi_2$} & $99.54\%$ & $100.00\%$ & $100.00\%$ & $98.20\%$\\
 \hline
& & & (140~m, 180~m) & (150~m, 190~m) & (160~m, 200~m) & (180~m, 220~m)\\
 \cline{2-7}
 \multirow{8}{*}{(100, 140)} & \multirow{4}{*}{4} & \multirow{1}{*}{HeBF} & $1.7975$ & $1.6442$ & $1.5096$ & $1.9727$\\
 \cline{3-7}
 & & \multirow{1}{*}{USRMNet} & $1.7922$ & $1.6364$ & $1.5069$ & $1.9691$ \\
 \cline{3-7}
 & & \multirow{1}{*}{$\varpi_1$} & $99.71\%$ & $99.53\%$ & $99.82\%$ & $99.82\%$ \\
 \cline{3-7}
 & & \multirow{1}{*}{$\varpi_2$} & $100.00\%$ & $99.96\%$ & $100.00\%$ & $100.00\%$ \\
 \cline{2-7}
  & \multirow{4}{*}{6} & \multirow{1}{*}{HeBF} & $1.4172$ & $1.2816$ & $1.1579$ & $0.9417$ \\
 \cline{3-7}
 & & \multirow{1}{*}{USRMNet} & $1.4064$ & $1.2702$ & $1.1460$ & $0.9293$ \\
 \cline{3-7}
 & & \multirow{1}{*}{$\varpi_1$} & $99.24\%$ & $99.83\%$ & $98.97\%$ & $98.68\%$ \\
 \cline{3-7}
 & & \multirow{1}{*}{$\varpi_2$} & $100.00\%$ & $100.00\%$ & $100.00\%$ & $99.82\%$ \\
 \hline
    \end{tabular}}
 \end{threeparttable}
\end{table*}

\subsection*{D. Computational Complexity Analysis}
In this subsection, we focus on analyzing the computational complexity of the USRMNet model. Then, we compare the USRMNet model with HeBF in terms of the computational complexity. It is worth noting that this computational complexity is an order of magnitude representation of time complexity, and the real computational complexity will be greater than it. The computational complexity of solving problem~\eqref{Cachenable09} in HeBF has been analyzed in~\cite{he2021beamforming}, which is $\mathcal{O}\left(\digamma\left(\left(K^2+3K\right)^{3.5}+N_{\mathrm{t}}^{2.7}\right)\right)$, where $\digamma$ is the update times of beamforming vector. While the computational complexity of the beamforming vector update~\eqref{Cachenable08} is $\mathcal{O}\left(KN_{\mathrm{t}}^3\right)$. Hence, the total computational complexity of HeBF is $\mathcal{O}\left(\digamma\left(\left(K^2+3K\right)^{3.5}+N_{\mathrm{t}}^{2.7}+KN_{\mathrm{t}}^3\right)\right)$. For the USRMNet model, the computational complexity of step 2 in Algorithm 1 is $K^2N_{\mathrm{t}}+\sum_{f=1}^{F_{\mathcal{N}_{\eta}}}\left(K^3\left(K_{\mathcal{N}_{\eta}}-1\right)+K^2K_{\mathcal{N}_{\eta}}+K^2p_{\mathcal{N}_{\eta}}^f\right)$, where $F_{\mathcal{N}_{\eta}}$, $K_{\mathcal{N}_{\eta}}$, and $p_{\mathcal{N}_{\eta}}^f$ are the number of graph convolutional layers, the highest power, immediate feature dimension of the $f$-th graph convolutional layer for $\mathcal{N}_{\eta}\left(\mathbf{\Theta}_{\eta};\mathbf{H},\mathbf{w},\mathbf{x}\right)$, respectively.  The computational complexity of step 3 in algorithm 1 is $M$. The computational complexity of step 4 in algorithm 1 is $K^2N_{\mathrm{t}}+\sum_{f=1}^{F_{\mathcal{N}_{x}}}\left(K^3\left(K_{\mathcal{N}_{x}}-1\right)+K^2K_{\mathcal{N}_{x}}+K^2p_{\mathcal{N}_{x}}^f\right)+MN_h$, where $F_{\mathcal{N}_{x}}$, $K_{\mathcal{N}_{x}}$, and $p_{\mathcal{N}_{x}}^f$ are the number of graph convolutional layers, the highest power, immediate feature dimension of the $f$-th graph convolutional layer for $\mathcal{N}_{x}\left(\mathbf{\Theta}_{x};\mathbf{H},\mathbf{w},\hat{\mathbf{x}}\right)$, respectively. Therefore, the total computational complexity of the USRMNet model is $L[KN_{\mathrm{t}}^3 + 2K^2N_{\mathrm{t}}+\sum_{f=1}^{F_{\mathcal{N}_{\eta}}}\left(K^3\left(K_{\mathcal{N}_{\eta}}-1\right)+K^2K_{\mathcal{N}_{\eta}}+K^2p_{\mathcal{N}_{\eta}}^f\right)+\sum_{f=1}^{F_{\mathcal{N}_{x}}}\left(K^3\left(K_{\mathcal{N}_{x}}-1\right)+K^2K_{\mathcal{N}_{x}}+K^2p_{\mathcal{N}_{x}}^f\right)+MN_h
+M]$. According to the simulation configurations, we set $L=2, N_{\mathrm{t}}=32, F_{\mathcal{N}_{\eta}}=F_{\mathcal{N}_{x}}=3, K_{\mathcal{N}_{\eta}}=K_{\mathcal{N}_{x}}=1$. The corresponding $\{p_{\mathcal{N}_{\eta}}^f\}_{f=1}^{3}$ and $\{p_{\mathcal{N}_{x}}^f\}_{f=1}^{3}$ are set as $\{5, 32, 2\}$ and $\{K+5, 32, 5\}$, respectively. We further assume $\digamma=3$. The computational complexity comparison results between the USRMNet model and HeBF are shown in Table~\uppercase\expandafter{\romannumeral3}. The ratio $\varpi_3$ is defined as $\varpi_3=\frac{\text{Computational complexity of USRMNet}}{\text{Computational complexity of HeBF}}\%$. It is not hard to find that the computational complexity of the USRMNet model becomes more and more competitive as $K$ increases. The reason for this is that the computational complexity of beamforming vector update is dominant when $K$ is small. While for the same antenna configuration, the computational complexity of beamforming vector update will not change, so the computational complexity of solving problem~\eqref{Cachenable09} in HeBF accounts for an increasing proportion. However, $K$ has less effect on the computational complexity of the USRMNet model than the procedure of solving problem~\eqref{Cachenable09} in HeBF, so the ratio $\varpi_3$ is decrease sharply with $K$ increases.


\begin{table} [htbp]
  \renewcommand{\tablename}
  \caption{\centering{ TABLE \uppercase\expandafter{\romannumeral3}} \protect \\ \quad\quad\quad\quad COMPUTATIONAL COMPLEXITY COMPARISON BETWEEN USRMNet AND HeBF}
  \\ \\
  \centering
  \begin{threeparttable}[b]
  \setlength{\tabcolsep}{2.5mm}{
    \begin{tabular}{| c | c | c | c | c |} \hline
 \multirow{2}{*}{\makecell[c]{Methods}} & \multicolumn{4}{c|}{$K$} \\ \cline{2-5}
 & 4 & 6 & 8 & 10  \\
 \hline
 \multirow{1}{*}{HeBF} & $7.77\times 10^5$ & $4.10\times 10^6$ & $2.00\times 10^7$ & $7.63\times 10^7$  \\
 \hline
 \multirow{1}{*}{USRMNet} & $2.68\times 10^5$ & $4.20\times 10^5$ & $5.48\times 10^5$ & $6.93\times 10^5$  \\
 \hline
 \multirow{1}{*}{$\varpi_3$} & $34.52\%$ & $10.24\%$ & $2.74\%$ & $0.91\%$  \\
 \hline
    \end{tabular}}
 \end{threeparttable}
\end{table}


\section*{\sc \uppercase\expandafter{\romannumeral5}. Conclusions}
In this paper, we propose an universal framework, i.e., UPGDNet, for a family of constrained optimization problems in wireless networks, which is designed based on PGD. Specifically, we firstly separate the constraints into two categories according to the coupling relations among optimization variables and the convexity of constraints. One category of constraints includes convex constraints with decoupling among optimization variables, and the other category of constraints includes non-convex or convex constraints with coupling among optimization variables. Then, for one category of constraints, we directly project them onto feasibility region, while using a neural network to project another category of constraints onto feasibility region. To train the UPGDNet, we also design a Lagrange primal dual learning framework with a multi-task objective function in an unsupervised manner. To verify the effectiveness of the UPGDNet, we utilize it to design a unrolling model, i.e., USRMNet, to solve the WSRMax problem in the scenario of multiuser uRLLC with finite blocklength transmission. Numerical results show that the UPGDNet can be trained efficiently using our proposed training scheme, and the USRMNet model has a comparable performance with the baseline algorithm on the basis of ensuring low computational complexity. In addition, the USRMNet model also has a acceptable generalization ability in terms of the user distribution.

\bibliographystyle{IEEEtran}
\bibliography{reference}

\end{document}